\documentclass[reprint,prb,10pt,superscriptaddress,twocolumn,floatfix]{revtex4-1}

\usepackage{amsmath}    % need for subequations
\usepackage{graphicx}   % need for pdf figures
\usepackage{epstopdf}	% need for eps figures
\usepackage{verbatim}   % useful for program listings
\usepackage{color}      % use if color is used in text
\usepackage{subfigure}  % use for side-by-side figures
\usepackage{hyperref}   % use for hypertext links, including those to external documents and URLs
\newcommand{\bo}{\boldsymbol}

\newcommand{\overbar}[1]{\mkern 1.5mu\overline{\mkern-1.5mu#1\mkern-1.5mu}\mkern 1.5mu}

\begin{document}
\label{qcalc_standalone}

\title{Orbital Josephson Interference in a Nanowire Proximity Effect Junction}
\author{Kaveh Gharavi}
\affiliation{Institute for Quantum Computing, University of Waterloo, Waterloo, Ontario N2L 3G1, Canada}
\affiliation{Department of Physics and Astronomy, University of Waterloo, Waterloo, Ontario N2L 3G1, Canada}
\author{Jonathan Baugh}
\email{baugh@uwaterloo.ca}
\affiliation{Institute for Quantum Computing, University of Waterloo, Waterloo, Ontario N2L 3G1, Canada}
\affiliation{Department of Physics and Astronomy, University of Waterloo, Waterloo, Ontario N2L 3G1, Canada}
\affiliation{Department of Chemistry, University of Waterloo, Waterloo, Ontario N2L 3G1, Canada}

\date{\today}

\begin{abstract}
A semiconductor nanowire based superconductor-normal-superconductor ($SNS$) junction is modeled theoretically. A magnetic field is applied along the nanowire axis, parallel to the current. The Bogoliubov-de Gennes equations for Andreev bound states are solved while considering the electronic subbands due to radial confinement in the $N$-section. The energy-versus-phase curves of the Andreev bound states shift in phase as the $N$-section quasiparticles with orbital angular momentum couple to the axial field. A similar phase shift is observed in the continuum current of the junction. The quantum mechanical result is shown to reduce to an intuitive, semi-classical model when the Andreev approximation holds. Numerical calculations of the critical current versus axial field reveal flux-aperiodic oscillations that we identify as a novel form of Josephson interference due to this orbital subband effect. This behavior is studied as a function of junction length and chemical potential. Finally, we discuss extensions to the model that may be useful for describing realistic devices. 

\end{abstract}
\maketitle
The Josephson effect is characterized by a current-phase relationship (CPR) linking macroscopic current flow to the phase gradient of the superconducting order parameter \cite{Josephson}. The precise form of the CPR for a superconducting weak link depends on intrinsic factors such as junction geometry, material properties, coherence lengths, etc., in addition to extrinsic variables like temperature and magnetic field. In superconductor-normal-superconductor ($SNS$) junctions in which the $N$-section is long enough to suppress direct tunnelling of Cooper pairs, but shorter than the $N$-section phase coherence length, a supercurrent may be carried by quasiparticles undergoing Andreev reflection at the $S$-$N$ interfaces \cite{andreev_orig_0,andreev_orig_1, prox_andreev,thouless_energy_1}. Planar $SNS$ junctions of width large compared to the $S$-section superconducting coherence length have been studied in great detail \cite{Barone1} (width refers to the dimension perpendicular to the current). These have revealed, for example, Fraunhofer oscillations of the critical current $I_c$ with respect to an externally applied out-of-plane magnetic field \cite{chiodi2012,sns1,sns2}. For junction widths comparable to the $S$-section coherence length, i.e. the narrow junction limit, this becomes a quasi-Gaussian, monotonic decay of the critical current \cite{chiodi2012,narrow_junction_theory_01,crouzy2013}. Recently, attention has been given to nanoscale, quasi one-dimensional (1D) $SNS$ junctions, such as those readily engineered by contacting semiconductor nanowires with superconducting leads \cite{Doh, proximityEffectInN,proximityEffectInAs,nilsson_InSb_junc,gul_core_shell_junc}. Gating the semiconducting $N$-section allows for modulating the supercurrent by controlling the chemical potential \cite{Doh, nilsson_InSb_junc}. The oscillations of the magnetoresistance of a nanowire $SNS$ junction in the voltage-biased state (i.e. no dc supercurrent) versus an axial magnetic field have been studied \cite{gul_core_shell_junc}. Efforts to realize Majorana fermion quasiparticles in 1D semiconductors with strong spin-orbit interaction and proximity coupling to a superconductor \cite{lutchynPRL2010_theory,MourikSci12,DengLundObs12,Das2012_1,DengLundObs12} have further raised interest in this type of junction. Theoretical results have indicated that the behaviour of the critical current in such a junction versus magnetic field and chemical potential can be used to identify topological phases \cite{AguadoSCreadout}.\\
\indent Previous theoretical descriptions of quasi-1D $SNS$ junctions \cite{narrow_junction_theory_01,narrow_junction_theory_02,Hammer_non-ideal-interface,crouzy2013} have not fully considered the effects of nanoscale confinement on the CPR, in particular the implications of orbital angular momentum coupling to an external magnetic flux.  Here we provide a quantum mechanical description of an idealized junction with a flux applied along the nanowire axis (parallel to the current). For a planar junction, no significant modification of the CPR with an axial flux is expected, as azimuthal motion of the carriers is absent. However, for a cylindrical geometry, azimuthal motion leads to a non-trivial effect which we identify as a previously unstudied form of Josephson interference. This is due to the coupling between Andreev quasiparticles (bound states and continuum states) with orbital angular momentum and the axial flux, which results in phase shifts of the energy-versus-phase for these current carrying states. The total current summed over all channels (occupied orbitals) can display interference. In contrast to Fraunhofer interference in wide planar junctions, the flux is aligned \emph{with} the current and the oscillations are \emph{not} periodic in the flux quantum. This effect is only present in nanoscale junctions with lateral dimensions (i.e. diameter) smaller than the London penetration depth. This is a regime in which the general theorem of Byers and Yang \cite{BY_theorem} does not apply. It is shown that the supercurrent from continuum states also contributes to this interference. For certain junction parameters, the interference effect can dominate the $I_c$ vs $\Phi$ characteristics. Semiclassically, the effect is intuitively understood by the pickup of a magnetic phase by Andreev pairs with an azimuthal velocity component as they cross the junction ballistically. The aim of this paper is to theoretically describe this type of Josephson interference in a fully quantum mechanical way. In particular, we are interested in understanding the effect in isolation from the additional complications of real devices, such as non-cylindrical contact geometry, interfacial potential barriers, etc. We consider in the discussion section how to modify the present model to better describe realistic devices. Here, we consider the case where the diameter is smaller than the superconducting coherence length in the $S$-section, so that the phase of the order parameter is uniform around the $S$-section circumference in any magnetic field up to the critical field of the leads, $H_c$. Spin-orbit and Zeeman effects in the $N$-section (e.g. relevant to III-V semiconductor nanowires) are neglected, and we assume no barriers at the $S$-$N$ interfaces. Furthermore, we neglect magnetic depairing effects.  
\section{Model}
\label{sec:1_model}
\begin{figure}[t!]
\label{fig:fig1_schematic}
\begin{center}
		\includegraphics[width=3.4in]{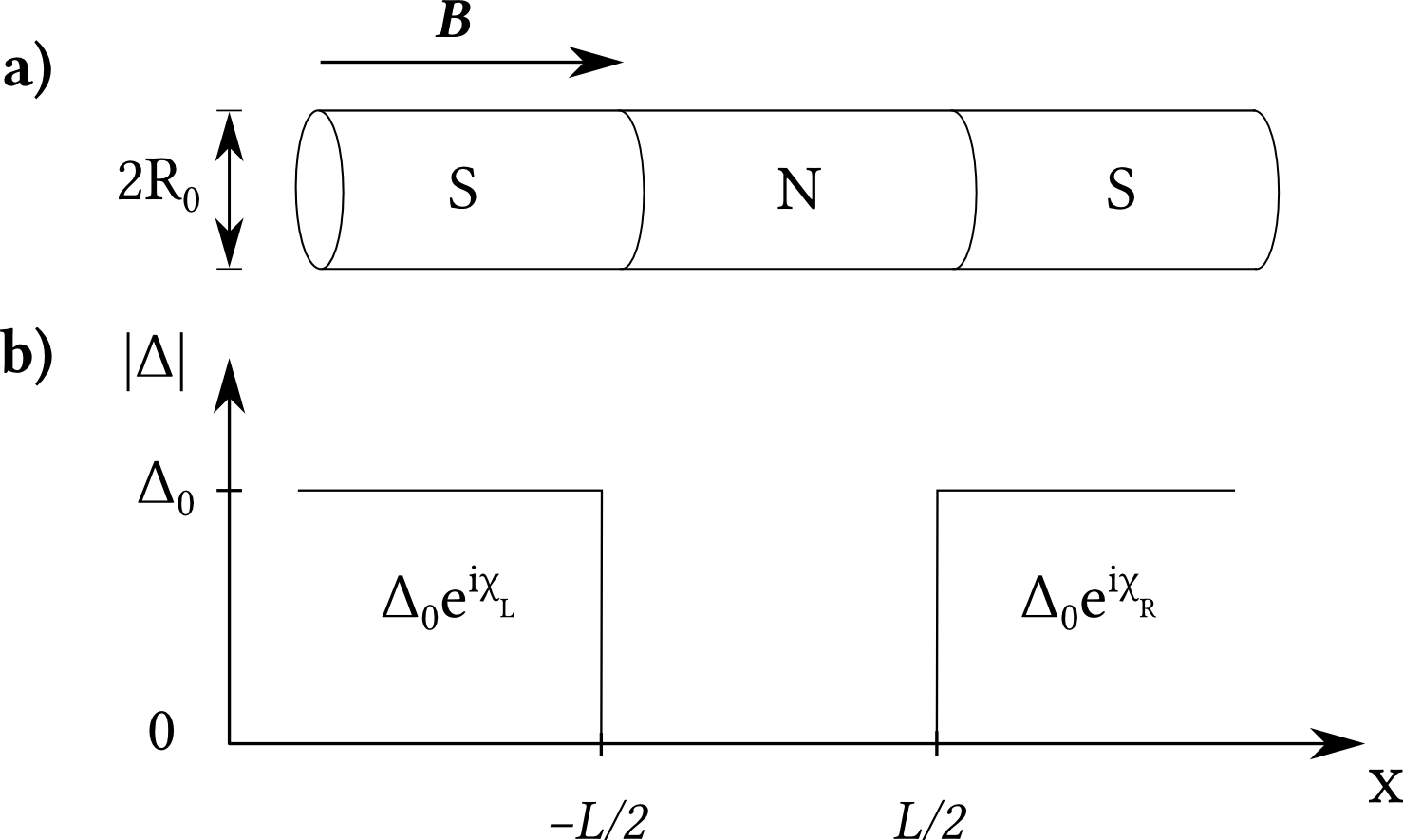}
	\end{center}
	\caption{a) Schematic of the nanowire $SNS$ junction of length $L$. It is modelled as a cylinder with a nanoscale diameter $d$ smaller than both the London penetration length and the $S$-section phase coherence length. An axial magnetic field $\bo B = B_\parallel \hat {\bo x}$ penetrates the cylinder. b) The superconducting order parameter has the magnitude $\Delta_0$ in the $S$-section, and is zero in the normal section, with a jumplike variation at the boundaries.}
\end{figure}
\begin{figure}[h!]
	\begin{center}
		\includegraphics[width=3.1in]{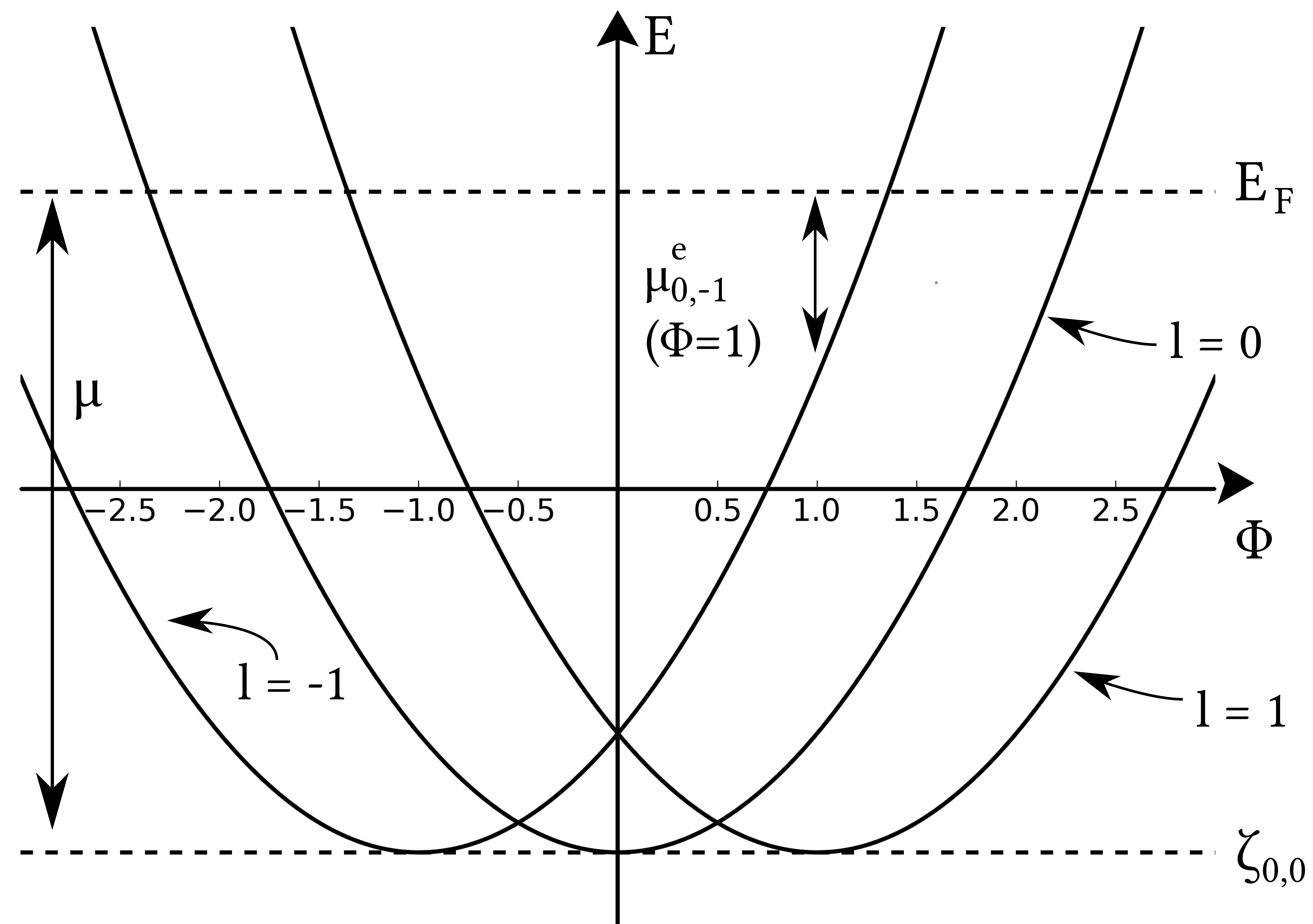}
	\end{center}
	\caption{Subband energies for the electrons in the $N$-section (i.e. the eigenenergies of $H_0$ in Eq. \ref{eq:h0} at $k=0$) versus the normalized magnetic flux $\Phi = (\pi B_\parallel R^2)/(h/e)$. The $l=0$ and $l=\pm 1$ subbands with $n=0$ are shown. The energies are parabolic because of the $\frac{\hbar^2}{2m^* R^2} (l - \Phi)^2$ contribution to the energy by $H_\theta$ (see Eqs. \ref{eq:h0}c, \ref{eq:single-particle-energy}a). The upper dashed line is the Fermi energy. The chemical potential $\mu$ is the Fermi energy measured from the bottom of the lowest subband, $\zeta_{0,0}$ (lower dashed line). We have assumed $\zeta_{0,\pm 1} = \zeta_{0,0}$, so the bottoms of the $n=0$ subbands have the same energy.  The effective chemical potential for electrons at flux $\Phi$, $\mu^e_{n,l} (\Phi)$, is the difference between the subband energy and the Fermi energy at that flux. This is shown for the subband $(n, l) = (0, -1)$ at $\Phi = 1$. For the hole-like states (not shown), the subband energies (Eq. \ref{eq:single-particle-energy}b) are inverted (mirrored) with respect to the $E = E_F$ line, and the $l$ quantum number negated ($l \rightarrow -l$). The effective chemical potential for holes is given by $\mu^h_{n,l}(\Phi) = \mu^e_{n,-l}(\Phi)$.}
	\label{fig:fig2_subbands}
\end{figure}

\indent Consider an $SNS$ junction created by a semiconducting nanowire contacted by superconducting leads. A cylindrical coordinate system $\bo r = (x, \rho, \theta)$ is used, with the nanowire axis along $\hat {\bo x}$. The junction is modelled as a cylinder of radius $R_0$. The diameter $d = 2R_0$ of the cylinder is assumed smaller than the $S$-section London penetration depth $\lambda_S$, and the $S$-section superconducting coherence length $\xi_S$. In figure~\ref{fig:fig1_schematic}a, we divide the cylinder into three regions, with region 1 the superconducting section corresponding to the left lead ($ x <- L/2$), region 2 the normal section corresponding to the nanowire ($\left| x \right|<L/2$),  and region 3 the $S$-section corresponding to the right lead ($ x > L/2$). The leads are connected to bulk superconductors at temperature $T$. Figure~\ref{fig:fig1_schematic}b shows the corresponding variation of the superconducting order parameter inside the nanowire.\\ 
\indent We assume uniform electrostatic potentials in each section (i.e. no scattering potential is included and transport is ballistic), and no potential barrier at the $S$-$N$ interfaces. The effective mass of the electron is assumed to have the same value $m^*$ in both $S$- and $N$-sections so that there is no Fermi wavevector mismatch (FWVM) at the $S$-$N$ interfaces. This assumption allows us to use Kulik's method \cite{kulik} to calculate the supercurrent of the junction. The advantage of this method is that it gives an analytical expression for the bound state energies, which provides us an intuitive way to understand the interference effect and connects our model to an approximate semiclassical picture. We note that inclusion of FWVM or interfacial barriers would not alter the basic mechanism of orbital Josephson interference that is demonstrated by this simpler model, but would require use of a more complicated transmission matrix formalism to calculate supercurrents. Our expectations for the qualitative effects of barriers are discussed in section \ref{sec:4_disc}.\\
\indent An axial magnetic field $\bo B = B_\parallel \bo{\hat{x}}$ penetrates the cylinder. Any screening of the magnetic field in the $S$-sections is neglected, as we have $d < \lambda_S$. In the Coulomb gauge, the vector potential is $\bo A = A_\theta  \bo{\hat{\theta}} = (B_\parallel \rho /2) \bo{\hat{\theta}}$. Using the superscript $\alpha = 1,2,3$ to refer to the three sections of the junction (with the $N$-section corresponding to $\alpha=2$), the single-electron Hamiltonian (excluding the superconducting pairing potential) in the presence of the magnetic field can be written as:
\begin{subequations}
	\label{eq:h0}
	\begin{align}
		H_0 &= - \mu + H_x + H_\theta + \sum_{\alpha = 1,2,3}V^\alpha(\rho);\\
		H_x &= -\frac{\hbar ^2}{2m^*}\frac{\partial ^2}{\partial x^2},\\
		H_\theta &=\frac{1}{2m^*}(-i \hbar \frac{1}{\rho} \frac{\partial}{\partial \theta} - e A_\theta)^2. \end{align}
\end{subequations}
Here, $H_x$ describes the kinetic energy of motion along the axis of the cylinder, $H_\theta$ the kinetic and magnetic energies of the azimuthal motion around the cylinder, and $V^\alpha(\rho)$ the radial confining potential of the cylinder in section $\alpha$.\\
\indent Radial confinement results in charge carriers occupying transverse subbands denoted by a pair of quantum numbers. We use the pair $(n,l)$ in the $N$-section, and $(p,l')$ in the $S$-sections, where $n$ and $p$ are the radial quantum numbers, and $l,l'$ the orbital angular momentum quantum numbers. The chemical potential in the cylinder is defined as the energy difference between the bottom edge of the lowest subband and the Fermi energy, and is denoted by $\mu$ (figure \ref{fig:fig2_subbands}). The numerical calculations were performed using the electron effective mass $m^* = 0.023 m_e$ corresponding to InAs. Zeeman and spin-orbit effects on the critical current of the junction (studied in Ref. \cite{yokoyama_spin_orbit_zeeman}) are not considered here, in order to focus on the effects of orbital angular momentum. In section \ref{sec:4_disc}, we discuss the conditions under which either the orbital effect or the Zeeman + spin orbit effects could be more dominant. \\
\indent In this paper, we do not write out an explicit form for $V^\alpha(\rho)$, and do not solve for the radial wavefunctions corresponding to the subbands $(n,l)$ in any section of the cylinder. Instead, we use a shell conduction model for the $N$-section. This is appropriate for certain III-V nanowires (such as InAs or InN), where the charge carriers are typically confined near the surface due to a positive surface potential, forming a surface accumulation layer \cite{inasAcc0, InN_accum}. Assuming a strong downward surface band bending ($\sim 100 - 200$ meV) \cite{band_bending_meas}, the radial position of the carriers in all subbands $(n,l)$ is taken to be $R \lesssim R_0$. This greatly simplifies the calculation of the eigenvalues of $H_\theta$ (Eq. \ref{eq:h0}c). However, we emphasize that the qualitative results obtained here should not be limited to this shell conduction model, particularly since we find a weak dependence of the interference effect on $R$. \\
\indent Superconductivity in the leads is described by the order parameter (pairing potential) $\Delta (\bo r)$, which in the general case, must be calculated self-consistently. We use a simplified model in which $\Delta$ is constant, so that proximity effects such as the reduction of $\Delta$ near the $S$-$N$ interfaces due to `reverse' proximity are neglected. For $\rho < R_0$, there is a jumplike variation at the boundary of each section (figure \ref{fig:fig1_schematic}b):
\begin{equation}
	\label{eq:delta}
	\Delta (\bo r) = \begin{cases}
			\Delta_0 e^{i \chi_L}, & \left. \, x \; \right. < -L/2\\
			0, & \left| x \right| < L/2 \\
			\Delta_0 e^{i \chi_R}. & \left. \, x \; \right. >L/2 \end{cases}
\end{equation}
Outside the radius of the cylinder, the order parameter is zero: $\Delta(\bo r) = 0$ when $\rho > R_0$. Here, $\Delta_0$ is the superconducting energy gap value in the leads, and $\chi_{L(R)}$ is the phase of the superconducting condensate in the left (right) lead.  The order parameter is zero in the $N$-section because of the lack of attractive electron-electron interactions (repulsive interactions are present in general, but neglected here.)\\
\indent In Eq. \ref{eq:delta}, a spatially uniform $\Delta$ is assumed in the $S$-sections in all magnetic fields up to $H_c$. This is justified if the following two conditions hold: (i) the diameter of the cylinder is smaller than the superconducting coherence length in the $S$-section, $ d < \xi_S$. The change in the phase of the order parameter around the circumference of the cylinder, $\delta \chi$, is constrained to integer multiples of $2 \pi$, because $\Delta$ has to be single valued (i.e. due to fluxoid quantization \cite{tinkham}). When $d < \xi_S$, one can assume $\delta \chi = 0$, since $\xi_S$ sets the length-scale for the spatial variations of the order parameter \cite{deGennes_book}, so $\Delta$ must be uniform. The validity of the assumption $d < \xi_S$ in experimental devices is discussed in section \ref{sec:4_disc}. (ii) The injected current in the $SNS$ junction is much smaller than the critical current of the superconducting leads. Otherwise, the superfluid flow in the $S$-sections cannot be neglected, and a self-consistent determination of $\Delta$ is required, as performed in Ref. \cite{bagwell_one_dim_2}. Here, we assume the critical current of the junction is bottle-necked in the $N$-section, which is reasonable given that the critical currents of nanowire junctions are typically small compared to those of the $S$ leads. 
\section{Theory}
\label{sec:2_theory}
We wish to calculate the current-phase relationship (CPR) of the junction in the presence of an axial magnetic field. First, the spectrum of discrete levels (Andreev bound states) in the junction is obtained. Next, the current from the discrete levels, as well the ``continuum" levels with energy  $|E| > \Delta_0$ is calculated. It is shown that energy eigenstates corresponding to these energy levels (both the bound states and the continuum states) follow the single-electron subband structure imposed by the Hamiltonian $H_0$ (Eq. \ref{eq:h0}). In particular, we show that the CPR is modified by the axial magnetic flux in a way that depends on the orbital angular momentum of the subbands. This leads to a form of Josephson interference when one or more subbands with orbital angular momentum are occupied.
\subsection{Bogoliubov-de Gennes equations}
\label{sec:sub2-1_BdG}
The wavefunctions of the elementary excitations of the $SNS$ junction are identified as the solutions to the Bogoliubov-de Gennes \cite{deGennes_book} (BdG) equations:
\begin{equation}\label{eq:bdg}
	\left( \begin{array}{cc}
		H_{0} & \Delta(\bo r)\\
		\Delta^*(\bo r) & -H^*_{0} \end{array}
		\right) \left( \begin{array}{c}
			u(\bo r) \\
			v(\bo r) \end{array}
			\right) = E \left( \begin{array}{c}
				u(\bo r)\\
				v(\bo r) \end{array}
				\right),
\end{equation}
where $H_0$ is given by Eq. \ref{eq:h0}, and $u(\bo r)$ and $v(\bo r)$ are particle- and hole-like wavefunctions. The asterisk ($^*$) denotes complex conjugation.\\
\indent The solution strategy for Eq. \ref{eq:bdg} starts with finding the solutions to $H_0$. Let us first consider the $N$-section.  Given the simple forms of $H_x$ and $H_\theta$ in Eq. \ref{eq:h0}, it is clear that the single-particle eigenfunctions are plane-waves in the $x, \theta$ directions of the form $e^{i k x} e^{i l \theta} \phi_{n,l}(\rho)$. The linear momentum along the axis of the cylinder is given by $\hbar k$, and the radial eigenfunction in the subband $(n,l)$ is given by $\phi_{n,l}(\rho)$. As discussed above, we use a shell conduction model for the $N$-section, so $\phi_{n,l}$ is not written out explicitly, but assumed to result in a radial position $R \lesssim R_0$ for the carriers. The most general solutions $u(\bo r), v(\bo r)$ to Eq. \ref{eq:bdg} are expansions over these single-particle solutions \cite{han-crespi,hollow_nanocylinder}. However, since $\Delta^*(\bo r)=0$ in the $N$-section, ($k,n,l$) are good quantum numbers and the single particle energies are given by: 
\begin{subequations}
	\label{eq:single-particle-energy}
	\begin {align}
		H_0 u(\bo r) = & \left\lbrace \hbar^2 k^ 2 / (2m^*) + \left[ \hbar^2/(2m^*R^2) \right] (l^2 + \Phi ^2) \right. \nonumber \\
		\;& \left. - \varepsilon_{l} +\zeta_{n,l} - \mu \right\rbrace u(\bo r),\\ 
		-H_0^* v(\bo r) = &  \left\lbrace- \hbar^2 k^ 2 / (2m^*) - \left[ \hbar^2/(2m^* R^2) \right] (l^2 + \Phi ^2) \right. \nonumber \\
		\;& \left. -\varepsilon_{l} - \zeta_{n,l} + \mu \right\rbrace v(\bo r). \end {align}
\end{subequations}
Here, $\Phi = (\pi B_\parallel R^2)/(h/e)$ is the normalized magnetic flux enclosed by the charge carriers, $\varepsilon_{l} = \left[ \hbar^2 /(2 m^* R^2)\right](2l \Phi)$, and $\zeta_{n,l}$ is the radial confinement energy associated with $\phi_{n,l}(\rho)$. The electron subband energies (i.e. the eigenvalues of $u$ at $k=0$) are plotted in figure \ref{fig:fig2_subbands}, and are parabolic in shape versus the magnetic flux. For the corresponding eigenvalues for $v$, the parabolas are inverted (mirrored with respect to the Fermi energy, $E_F$). This gives $v(\bo r)$ its hole-like character: its group velocity $\bo v_g = \frac{1}{\hbar} \bo \nabla_{\bo k} E$ is opposite to its wave vector $\bo k$ (i.e. it is retroreflected -- see Ref. \cite{schapers_book}). Note, however, that for a given $k$, charge is transported in the same direction by the two wavefunctions, as the retroreflected hole has opposite charge to the electron. The eigenvalues associated with $u(\bo r), v(\bo r)$ are not equal in magnitude, as the term $\varepsilon_{l}$ has the same sign in both lines of Eq. \ref{eq:single-particle-energy}. This follows because of the complex conjugation of the diagonal term on the second row of Eq. \ref{eq:bdg}, and is a manifestation of the breaking of time-reversal symmetry in the presence of a magnetic field: the retroreflected particle sees the same magnetic field as the incident particle, rather than a time-reversed field $A_\theta \rightarrow -A_\theta$.\\
\subsection{Andreev bound states}
\label{sec:sub2-2_bound-states}
Following the original work of Kulik \cite{kulik} we calculate the spectrum of the bound states of the nanowire $SNS$ junction, however, here we allow the solutions to carry finite orbital angular momentum.\\
\indent Suppose there is a solution $\Psi(\bo r) = \left( u(\bo r), v(\bo r) \right) ^T $ to Eq. \ref{eq:bdg}, with energy $E$ within the gap, $\left|E \right| < \Delta_0$. Since we assume no FWVM or barriers at the $S$-$N$ interface, the right- and left-moving solutions $\Psi^\pm$ can be separated \cite{kulik}. We disallow superpositions of $(n,l)$ subbands in the $N$-section. This is justified because: (i) we have assumed the ballistic regime, so no scattering-induced subband-mixing occurs, and (ii) the pairing potential, Eq. \ref{eq:delta}, is zero in the $N$-section, and so it does not mix the $(n,l)$ states (see Ref. \cite{hollow_nanocylinder}). \\
\indent In the $S$-sections, Cooper pairing generally mixes subbands with different radial quantum numbers $p$, but the orbital angular momentum number $l$ remains a good quantum number \cite{han-crespi}. The latter follows from the cylindrical symmetry of Eq. \ref{eq:bdg}, which in turn follows from a cylindrically symmetric $H_0$ and a spatially uniform $\Delta$. For a given quantum number $l$ and energy $E$, the most generic single-particle wavefunction in the leads is given by $e^{il\theta}\sum_p \beta_{p,E} e^{ik_{p,E} x}\phi_{p,l} = e^{il\theta} Y_{l,E} (x,\rho)$. In each term of the sum, $k_{p,E}$ adjusts itself such that the energy of that term is $E$. For $|E| < \Delta_0$ considered here, $k_{p,E}$ also has an imaginary component, resulting in an exponential decay of the wavefunction inside the $S$-section \cite{schapers_book}. In general, there is significant freedom in the choice of the expansion coefficients $\beta_{p,E}$ which will allow matching of the radial wavefunctions in the $N$ and $S$ sections. \\
\indent The wavefunctions $\Psi^\pm$ can be written as:
\begin{equation}
	\Psi^\pm_{n,l,E} = \begin{cases}
		A^\pm e^{il\theta} e^{\pm ik^e_{n,l}x} \phi_{n,l}(\rho) \left(  \begin {array}{c} 1 \\ 0 \end{array} \right) + & \; \\		B^\pm e^{il\theta} e^{\pm ik^h_{n,l}x} \phi_{n,l}(\rho) \left(  \begin {array}{c} 0 \\ 1 \end{array} \right), & \left| x \right| < L/2 \\
		C^\pm e^{il\theta} \psi^{R}_{l}(x,\rho) 	\left(  \begin {array}{c} e^{i\chi_R} \\ \gamma^\pm \end{array} \right),& x>L/2 \\
		D^\pm e^{il\theta} \psi^{L}_{l}(x,\rho) 	\left(  \begin {array}{c} \gamma^\pm \\ e^{-i\chi_L} \end{array} \right).& x< - L/2 \\
	\end{cases}
	\label{eq:psi_plus}
\end{equation}
The wavenumbers $k^e_{n,l}$ and $k^h_{n,l}$ represent the momenta of the electron-like and hole-like components in the $N$-section, respectively. $\gamma^+ = \Delta_0 \left( E + i \sqrt {\Delta_0^2 - E^2} \right)^{-1}$ is the BCS coherence factor in the leads, and $\gamma^-$ is its complex conjugate. $\psi_l^{L,(R)}$ are, in general, superpositions of the $Y_{l,E}$ functions with different $E$, due to the inter-subband mixing induced by $\Delta$. Since there is no FWVM or barrier at the $S$-$N$ interfaces, $\Psi^\pm$ has to be continuous at $|x| = L/2$. In order for this to be possible, we must have $\psi^R_l(L/2,\rho) = \psi^L_l(-L/2,\rho) = \phi_{n,l}(\rho)$. A solution can always be achieved by a correct choice of the expansion coefficients $\beta_{p,E}$, so the form given in Eq. \ref{eq:psi_plus} for $\Psi^{\pm}$ is valid.\\  
\indent We now concentrate on the $N$-section, and derive the quantization rules for the energies of the bound states. Asserting that each term of $\Psi^{\pm}$ in the $N$-section has energy $E$, the wavenumbers $k^e_{n,l}, k^h_{n,l}$ are obtained as a function of energy:
\begin{subequations}
	\label{eq:wavenumbers}
	\begin {align}
		k^e_{n,l}(E) = \frac{\sqrt{2m^*}}{\hbar} \sqrt{\mu^e_{n,l} +E}, \\
		k^h_{n,l}(E) = \frac{\sqrt{2m^*}}{\hbar} \sqrt{\mu^h_{n,l} -E}, \end{align} %\\
\end{subequations}
where we have defined an effective chemical potential for an electron-like (hole-like) particle in the subband $(n,l)$ in the $N$-section $\mu^{e(h)}_{n,l} := \mu - \frac{\hbar^2}{2m^* R^2} (l \mp \Phi)^2 - \zeta_{n,l}$. The minus (plus) sign in the parentheses refers to the electron-like (hole-like) particle, and $\zeta_{n,l}$ is the radial confinement energy due to $\phi_{n,l}$. The effective chemical potential is the difference between the energy of the subband $(n,l)$ and the Fermi energy at a given magnetic field (see figure \ref{fig:fig2_subbands}), and is a positive quantity for any subband that is occupied.\\
\indent The energy quantization rules can be obtained \cite{kulik} by finding the set of coefficients $\left\lbrace A^\pm, \ldots,D^\pm \right\rbrace$ that make Eq. \ref{eq:psi_plus} continuous at $|x| = L/2$. This is only possible if the following relation holds:
\begin{equation}
	\gamma^2 e^{i (k^e_{n,l} - k^h_{n,l})L} e^{\mp i\chi} = 1,
	\label{eq:quantization-rule}
\end{equation}
where the junction phase $\chi = \chi_R - \chi_L$ enters with a minus sign when for $\Psi^+$, and a plus sign for $\Psi^-$. Note that the superscript $s=+,-$ denotes the right- and left-moving solutions, respectively. The complex phase of the right-hand side of Eq. \ref{eq:quantization-rule} must equal $2 m \pi$, with $m = 0,\pm 1,\pm 2,$ etc. Since $k^e_{n,l}, k^h_{n,l}$ depend explicitly on the bound state energy $E$ (Eq. \ref{eq:wavenumbers}), this results in a quantization rule for $E$, and yields the bound state spectrum. This procedure is carried out in section \ref{sec:3_results} to numerically solve for $E$ as a function of $\chi$.\\
\indent Quite generally, if $\Psi^s = (u, v)^T$ is an eigensolution of the BdG equation (Eq. \ref{eq:bdg}) with energy $E$, then $\overbar{\Psi}^s = (-v^*, u^*)^T$ is also an eigensolution\cite{deGennes_book} with energy $-E$. If $\Psi^s$ is a right-moving solution, then $\overbar{\Psi}^s$ is left-moving, and vice versa. Let $\overbar s$ denote the conjugate of $s$. The wavefunctions $\overbar{\Psi}^s$ and $\Psi^{\overbar{s}}$ are degenerate at zero field, for all $\chi$ (figure \ref{fig:fig3_BS_phase_shift}a). This degeneracy is lifted in the presence of the magnetic field, which induces a finite phase shift, as will be discussed below. The pair of solutions ($\Psi^+, \overbar{\Psi}^+$) are phase shifted \textit {together} in one direction, while the opposite pair ($\Psi^-, \overbar{\Psi}^-$) are phase shifted in the \textit{opposite} direction (figure \ref{fig:fig3_BS_phase_shift}b).\\
\subsection{Andreev approximation}
\label{sec:subsubsec_andreev}
\begin{figure}[t!]
	\begin{center}
		\includegraphics[width=3.5in]{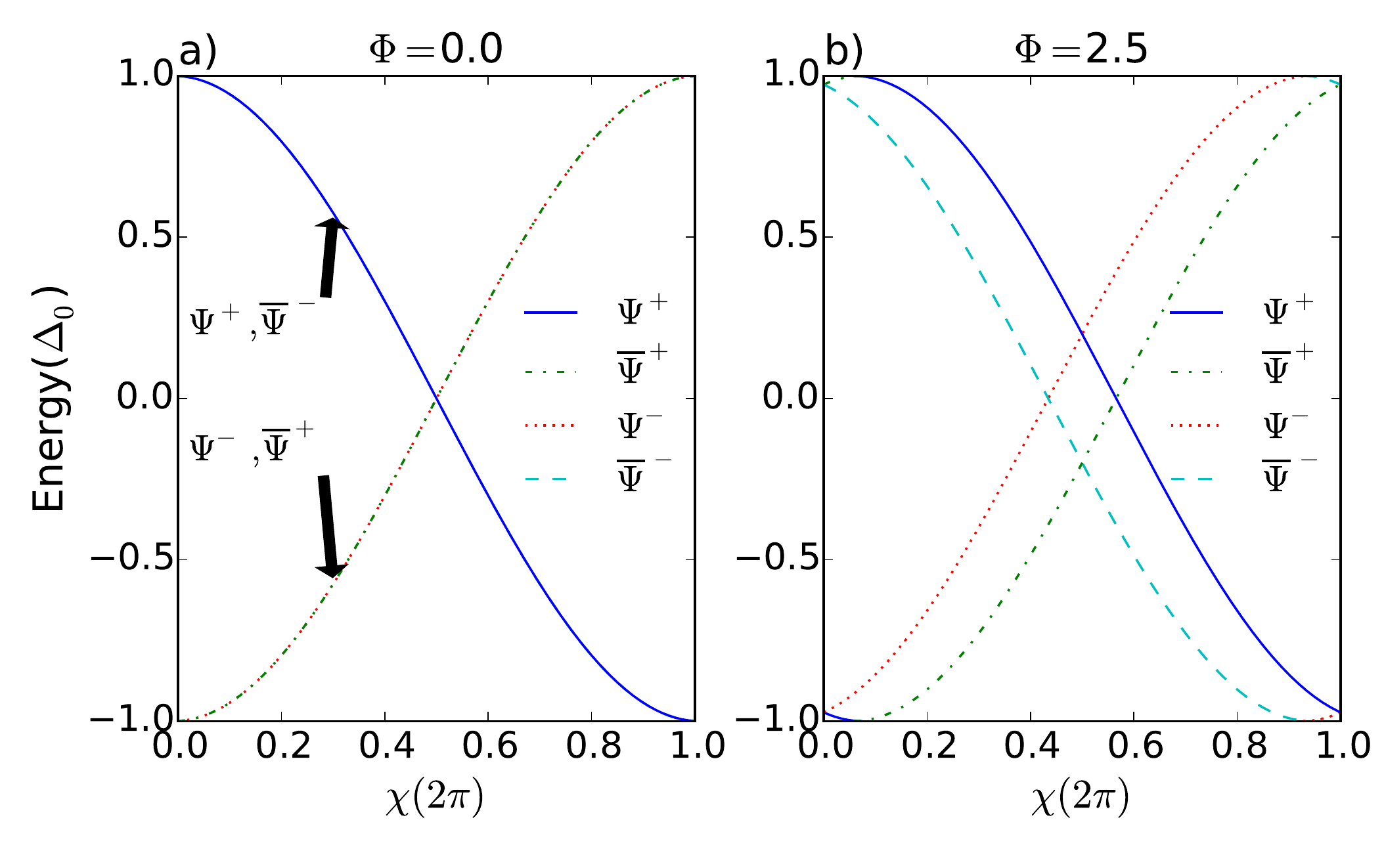}
	\end{center}
	\caption{Eigenenergies of the Andreev bound states of a short ($L = 25\; \mathrm{nm} \ll \xi^0_{n,l} \sim 200\; \mathrm{nm}$), cylindrical $SNS$ junction with no barriers at the $S$-$N$ interfaces, vs the superconducting phase difference $\chi$ of the leads. Two values of normalized magnetic flux, $\Phi = \pi R^2 B_\parallel / (h/e) = 0, 2.5$, are shown.  We concentrate on one subband $(n,l)$ with $n=0$ and $l=1$. a) Zero magnetic field, $\Phi = 0$. The energies correspond to the four allowed wavefunctions (defined in the main text), and the states are pairwise degenerate for all $\chi$. b) $\Phi = 2.5$, where the degeneracies are lifted, and there is a phase shift with opposite directions for each pair of states. The phase shift is small because of the short length of the junction. The following parameters were used for both panels: $\mu = 200 \; \mathrm{meV}, R = 30 \; \mathrm{nm}, T = 100\; \mathrm{mK}$.}
	\label{fig:fig3_BS_phase_shift}
\end{figure}

\indent Deriving an analytical expression for the bound state spectrum by inserting Eq. \ref{eq:wavenumbers} into Eq. \ref{eq:quantization-rule} becomes intractable, because of the complicated dependence of $k^e_{n,l} - k^h_{n,l}$ on $E$. In order to gain insight into the behaviour of the bound states, we invoke below the well-known Andreev approximation \cite{andreev_orig_0,andreev_orig_1,andreev_0}, in which $|k^e_{n,l} - k^h_{n,l}|$ is considered a small quantity compared to $|k^e_{n,l}|$ and $|k^h_{n,l}|$. This approximation is widely used in the literature for a variety of situations \cite{kulik,bagwell,andreev_0,andreev_1}, but can be violated in some regimes of our $SNS$ junction. In particular, when the subband energy is close to the Fermi energy, $k^e_{n,l}$ and $k^h_{n,l}$ become small and the assumption $|k^e_{n,l} - k^h_{n,l}| \ll |k^e_{n,l}|,|k^h_{n,l}|$ is not justified. Keeping these restrictions in mind, we now look at how the CPR of the junction is modified in the presence of the axial magnetic field.\\
\indent The effective chemical potential for electron-like (hole-like) particles in the subband $(n,l)$ in the $N$-section can be written $\mu^{e(h)}_{n,l}  = \mu - \frac{\hbar^2}{2m^* R^2} (l^2 + \Phi^2) \pm \varepsilon_{l} - \zeta_{n,l}$, where $\varepsilon_{l} = \left[ \hbar^2 / (2 m^* R^2) \right] (2l \Phi)$ enters with a plus sign for electron-like particles. It reflects the coupling of the orbital motion and the axial field. The Andreev approximation translates to the following condition: $E + \varepsilon_{l} \ll \mu^{e}, \mu^{h}$, i.e. the quasi-particle energy and the coupling to the field are small perturbations on the single-particle energies. Eq. \ref{eq:wavenumbers}a, \ref{eq:wavenumbers}b can be expanded in a Taylor series in the powers of $(E + \varepsilon_{l})$. We calculate $k^e_{n,l} - k^h_{n,l}$ to first order:
\begin{equation}
	\label{eq:wavenum-approx}
	k^e_{n,l} - k^h_{n,l} \simeq \frac{2}{\hbar} \frac{E + \varepsilon_{l}}{v_{n,l}},
\end{equation}
where $v_{n,l} = \sqrt{2 \left(\mu - \frac{\hbar^2}{2m^* R^2} (l^2 + \Phi^2) - \zeta_{n,l} \right) / m^*}$ is the velocity of a particle in the subband $(n,l)$ travelling along the cylinder axis in the $N$-section.\\
\indent By inserting $(k^e_{n,l} - k^h_{n,l})$ into Eq. \ref{eq:quantization-rule} and equating the complex phase of the left hand side of Eq. \ref{eq:quantization-rule} with $2 m \pi$, where $m = 0,1,2,$ etc., we obtain the following expression for the spectrum of bound states:
\begin{align}
	\left(\frac{L}{\xi^0_{n,l}} \right) \left( \frac{E^\pm_{n,l,m}}{\Delta_0} \right)& - 2\mathrm{arccos}\left( \frac{E^\pm_{n,l,m}}{\Delta_0} \right) \; \nonumber \\ \;& \mp \chi + \left( \frac{L}{\xi^0_{n,l}} \right) \left( \frac{\varepsilon_{l}}{\Delta_0} \right) = 2\pi m,		\label{eq:spectrum}
\end{align}
where $\xi^0_{n,l} = \hbar v_{n,l} / (2 \Delta_0)$ is the healing length \cite{bagwell} for the subband ($n,l$), and the energy of the bound state depends on three quantum numbers $n,l,m$, and the junction phase difference $\chi$. $E^+ (E^-)$ refers to the eigenenergy of the right-moving (left-moving) solution. Note that the energy corresponding to $\overbar{\Psi}^+$ is $-E^+$, for example. \\
\indent In a short junction, $L \ll \xi^0_{n,l}$, Eq. \ref{eq:spectrum} allows only one $m$ value per solution, and there are four bound states ($\Psi^+, \overbar{\Psi}^+, \Psi^-, \overbar{\Psi}^-$) per subband $(n,l)$. At zero field, there are two positive, and two negative solutions at any given $\chi$ (figure \ref{fig:fig3_BS_phase_shift}a). For long junctions there are more than four bound state energies per subband, with different $m$ numbers \cite{bagwell,bardeen_long_j,chrestin_sm_oscillations, schapers_book_1,Wees_long_j}.\\
\indent The bound state spectrum Eq. \ref{eq:spectrum} gives, for the case of no magnetic field ($\varepsilon_{l}$ = 0), a result similar to the well known Andreev levels of a ballistic $SNS$ junction \cite{kulik, bagwell}, but with a different value of the healing length for each subband. An example is shown in figure \ref{fig:fig3_BS_phase_shift}a for subband $(n,l) = (0,1)$ in a short junction with $L = 25\; \mathrm{nm} \ll \xi^0_{n,l} \sim 200\; \mathrm{nm}$. Figure \ref{fig:fig3_BS_phase_shift}b shows the energy-versus-phase curves of the bound levels at a finite flux, $\Phi = 2.5$. The curves are now phase shifted by an amount $\delta_{n,l} = \left( \frac{L}{\xi^0_{n,l}} \right) \left( \frac{\varepsilon_{l}}{\Delta_0} \right)$, where $\varepsilon_{l} = \left[ \hbar^2 /(2 m^* R^2)\right](2l \Phi)$. That is, $E^\pm_{n,l,m}(\chi) \rightarrow E^\pm_{n,l,m}(\chi \mp \delta_{n,l})$.
\subsection{Reduction to a semiclassical model}
\label{sec:semiclassical}
The phase shift $\delta_{n,l}$ can be understood semiclassically as the phase picked up by azimuthal travel around the circumference of the cylinder in the presence of the magnetic field. In this picture, for a subband $(n,l)$ with $l \neq 0$, the particles (both electron- and hole-like) travel in a spiral path as they traverse the junction length $L$. In the shell-conduction model the spiral has radius $R$. The velocity along the axis is $v_{n,l}$, while the azimuthal velocity is $v_\theta(l) = \hbar l / (m^* R)$. The semiclassical phase $\delta_{sc}$ is is due to the coupling of $v_\theta$ and the vector potential $\bo A = (B_\parallel \rho /2) \hat{\bo \theta}$, and is calculated from the Ginzburg-Landau formula for the phase:
\begin{equation}
\delta_{sc} = (2e/\hbar)\int \bo A \cdot \mathrm{d} \bo l= (2e/\hbar) \int \bo A \cdot \bo v \mathrm{d} t,
\end{equation}
where the differential element $\mathrm {d} \bo l$ is along the spiral path, $\bo v = v_{n,l} \hat{\bo x} + v_\theta \hat{\bo \theta}$ is the velocity, $t$ is time, and the second integral is taken from $t=0$ corresponding to the particle leaving one $S$-section, to $t = L/v_{n,l}$, when it arrives at the other $S$-section. The result is $\delta_{sc} = {e l L  B_\parallel}/({m^* v_{n,l}})$, which equals $\delta_{n,l}$. This shows that when the Andreev approximation holds and there is shell conduction, the semiclassical result coincides with the quantum mechanical one. Note that in the expression for $\delta_{sc}$ there is no explicit dependence on $R$. The dependence of the phase shift on $R$ comes only through $v_{n,l}$, and is a weak dependence within the Andreev approximation. More generally, we numerically calculate the energy spectra using Eq. \ref{eq:wavenumbers} without the Andreev approximation and find similar phase shifts that are always proportional to the junction length $L$ and to the angular momentum quantum number $l$. \\
\subsection{Bound state and continuum currents}
\label{sec:sub2-3_currents}
The current due to the Andreev bound states in the subband ($n,l$) at temperature T is calculated \cite{schapers_book_1, bagwell, beenakker_limit} from the formula
\begin{equation}
	I_{n,l}(\chi) = \frac {e}{\hbar} \sum_{s,m} f(E^s_{n,l,m}) \frac { \mathrm d E^s_{n,l,m}(\chi)}{\mathrm d \chi},
	\label{eq:current-subband}
\end{equation}
where $f(E^s_{n,l,m}) = 1/(\mathrm{exp}(E^s_{n,l,m}/(k_B T)) +1)$ is the Fermi-Dirac occupation probability of a given energy level ($k_B$ is the Boltzmann constant). Energies corresponding to both types of wavefunctions $\Psi^s_{n,l}$ and $\overbar{\Psi}^s_{n,l}$ must be inserted into Eq. \ref{eq:current-subband}. The total bound state current is the sum of supercurrent amplitudes from all occupied subbands \cite{schapers_book_2} (`open channels' in the language of Ref. \cite{Furusaki1992}):
\begin{equation}
	I_{\mathrm{total}}(\chi) = \sum_{n,l} I_{n,l}(\chi).
	\label{eq:current-total}
\end{equation}
\indent The continuous spectrum of states with energies $|E| > \Delta_0$ also contributes to the junction current.  A continuum level can be viewed as a ``leaky" solution \cite{bagwell,bardeen_long_j} to the Andreev bound state problem described above, with a complex-valued eigenenergy $E = E_R + iE_I$, with $|E_R| > \Delta_0$. The leaky level follows the same subband structure as the Andreev bound states. The imaginary component of energy results in a finite lifetime for the continuum level, reducing its contribution to the junction current, but for a long junction $L \gtrsim \xi^0_{n,l}$, this contribution is significant, and cannot be ignored \cite{bagwell}.\\
\indent The continuum current due to the subband $(n,l)$, $J_{n,l}(\chi)$, is calculated using the transmission formalism \cite{bagwell,Wees_long_j,tang}. We calculate the transmission coefficients for the electrical currents carried by electron-like and hole-like excitations incident on the $S$-$N$ interfaces, resulting in leaky solutions in the $N$-section. The details of the calculation are given in Appendix \ref{sec:appendix_cont}. For the rest of this section the subscripts $n,l$ are dropped for the sake of simplicity; it is implicitly assumed that all quantities pertain to the subband $(n,l)$. The result for the continuum current is:
 \begin{align}
J(\chi) = \frac{e}{h} \left( \int_{-\infty}^{-\Delta_0} + \int_{\Delta_0}^{\infty} \right) |u_0^2 - v_0^2| \left(\frac{1}{F^+(E,-\chi)} \right. \nonumber \\
 \left. - \frac{1}{F^-(E,-\chi)} - \frac{1}{F^+(E,+\chi)} + \frac{1}{F^-(E,+\chi)} \right) f(E) \mathrm{d} E , 
\label{eq:continuum-current}
\end{align}
where $E$ is the real part of the energy of the continuum level and $f(E)$ is the Fermi-Dirac distribution at temperature $T$, and $u_0, v_0$ are real-valued BCS coherence factors:
\begin{subequations}
\begin{align}
u_0^2 = \frac{1}{2}\left( 1+\sqrt{E^2 - \Delta_0^2}/E \right),\\
v_0^2 = \frac{1}{2}\left( 1-\sqrt{E^2 - \Delta_0^2}/E \right). \end{align}
\label{eq:coherence-factors}
\end{subequations}

\noindent The functions $F^\pm(E,\chi)$ depend on energy, through the wavenumbers $k^e$ and $k^h$ (Eq. \ref{eq:wavenumbers}) as well as the coherence factors $u_0, v_0$: $F^\pm(E,\chi)= u_0^4 + v_0^4 - 2 u_0^2 v_0^2 \mathrm{cos} \left[ (k^e(\pm E) - k^h(\pm E)) L + \chi \right]$. Equation \ref{eq:continuum-current} can be intuitively understood in terms of the leaky solutions to the BdG equation: the terms containing $F^+$ pertain to the contribution of leaky states of type $\Psi^s_{n,l,m}$, while those containing $F^-$ pertain to $\overbar \Psi^s_{n,l,m}$. The junction phase $\chi$ enters with a plus (minus) sign for left (right) moving solutions. At zero magnetic field, $F^+(E,\chi) = F^-(E,-\chi)$. This is analogous to the degeneracy of $\Psi^+, \overbar {\Psi}^-$ in figure \ref{fig:fig3_BS_phase_shift}a. Therefore, Eq. \ref{eq:continuum-current} reduces to Eq. 17 in Ref. \cite{bagwell}, up to an application of the Andreev approximation.\\
\indent In the presence of the magnetic field, the terms $(k^e - k^h)L$ shift the functions $F^\pm(E,\chi)$ in phase relative to the zero-field case, in the same manner as the phase shifts found previously for the bound states. Employing the Andreev approximation (Eq. \ref{eq:wavenum-approx}), we obtain
\begin{equation}
F^\pm(E,\chi) = u_0^4 + v_0^4 - 2 u_0^2 v_0^2 \mathrm{cos} \left[ \left(\frac{E \pm \varepsilon_{l}}{\Delta_0}\right)\left(\frac{L}{\xi^0}\right) + \chi \right],
\label{eq:D_cont_approx}
\end{equation}
which is shifted in phase with respect to the zero-field case. Explicitly for the subband $(n,l)$ we have $F_{n,l}^\pm(E,\chi) \rightarrow F_{n,l}^\pm(E, \chi \pm \delta_{n,l})$, with the phase shift $\delta_{n,l}$ defined previously.\\
\indent The total continuum current of the junction is 
\begin{equation}
J_{\mathrm{total}} (\chi) = \sum_{n,l} J_{n,l} (\chi).
\label{current-continuum-total}
\end{equation}
The critical current $I_c$ of the junction is defined as the maximum of total bound state + continuum currents with respect to $\chi$:
\begin{equation}
	I_c = \mathrm{max}_{\chi \in [0, 2 \pi) } \left[ I_{\mathrm{total}}(\chi) + J_{\mathrm{total}}(\chi) \right].
	\label{eq:current-critical}
\end{equation}
\section{Numerical Results}
\label{sec:3_results}
We numerically solve the continuum and bound state currents of an $SNS$ junction at finite magnetic fields, using the shell conduction approximation with the shell at a radius $R = 30\; \mathrm{nm}$. Temperature is set to $T = 100\; \mathrm{mK}$ in all calculations. From this point on, only $(n,l)$ subbands with $n=0$ are assumed to be occupied in the $N$-section. The Andreev approximation is not used in calculating the CPR. The critical current of the junction is calculated from Eq. \ref{eq:current-critical}, and its behaviour versus axial magnetic flux $\Phi = \pi R^2 B_\parallel / (h/e)$ is studied. Note that our assumption of no barriers at the $S$-$N$ interfaces also implies full Andreev reflection.
\subsection{Single subband}
\label{sec:31_one_subband}
\begin{figure}[t]
	\begin{center}
		\includegraphics[width=3.5in]{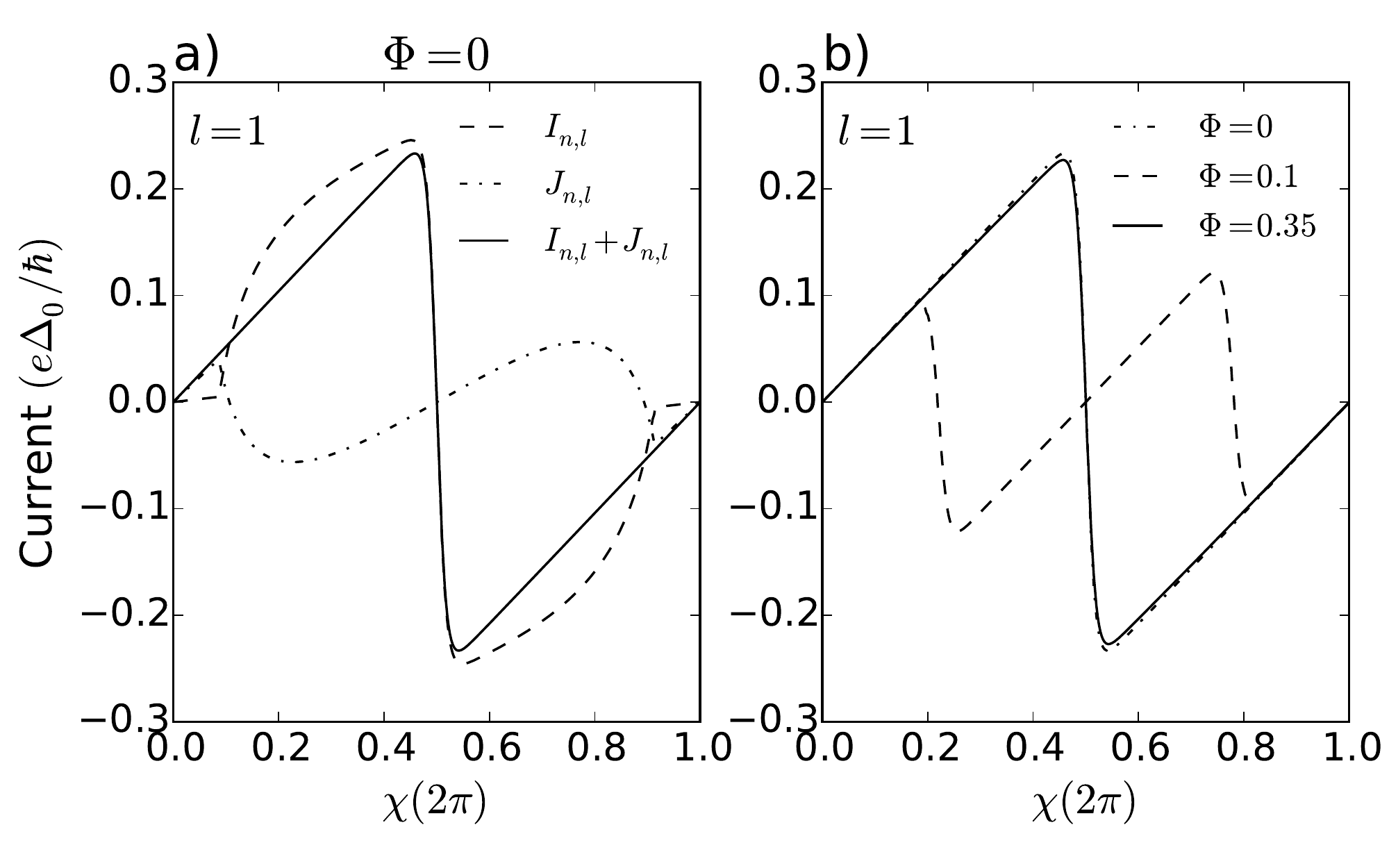}
	\end{center}
	\caption{a) Bound state current $I_{n,l}$, continuum current $J_{n,l}$, and the sum $I_{n,l}+ J_{n,l}$ for the subband $(n,l) = (0,1)$ vs the superconducting phase $\chi$ at zero magnetic field, $\Phi = 0$. Since the junction is long, $L = 500\; \mathrm{nm} \gtrsim \xi^0_{n,l}$, the CPR is triangular. The kinks in $I_{n,l}, J_{n,l}$ at $\chi = 0.05, 0.95$ are due to Andreev bound states crossing the gap edge into the continuum levels, but the total subband current is a smooth function of $\chi$. b) Total subband current $I_{n,l} + J_{n,l}$ vs the superconducting phase as a function of the normalized axial magnetic flux $\Phi$. At zero flux the maximal value occurs near $\chi = \pi$. At finite flux, the bound state and continuum currents are phase shifted. For $\Phi = 0.1$, two discontinuities can be seen in $I_{n,l} + J_{n,l}$ because of the phase shifts, and the maximal value no longer occurs near $\chi = \pi$. At $\Phi = 0.35$, the phase shifts amount to $2\pi$ and the zero-field curve is recovered. The maximal current at this flux is slightly smaller than the zero field case, due to a decrease in the average momentum of the Andreev quasiparticles with increasing flux. The following parameters were used in both panels: $L = 500\; \mathrm{nm}, \mu = 8.5 \; \mathrm{meV}, R = 30 \; \mathrm{nm}, T = 100\; \mathrm{mK}$.}
	\label{fig:fig4_currents}
\end{figure}
\indent As an illuminating example we study a 500 nm long junction with a chemical potential of $\mu = 8.5\; \mathrm{meV}$. This value for $\mu$ is chosen because it allows $l = -1,0,1$ subbands to be occupied at $\Phi = 0$; at $\Phi = 1$ the $|l| = 1$ subbands depopulate \footnote{The effective chemical potentials follow $\mu^e_{n,l} (\Phi) = \mu^h_{n,-l} (\Phi)$, see figure \ref{fig:fig2_subbands}. In this example, as $\Phi$ approaches 1, $\mu^e_{0,1}, \mu^h_{0,-1}$ go to zero, at which point Andreev quasiparticles can no longer be supported by the $|l|=1$ subbands.}. In this section we concentrate on the CPR obtained for one subband, namely $l = 1$, and discuss how the coupling of the finite angular momentum with the axial field modifies the subband CPR.

\begin{figure*}[t!]
	\begin{center}
		\includegraphics[width=7in]{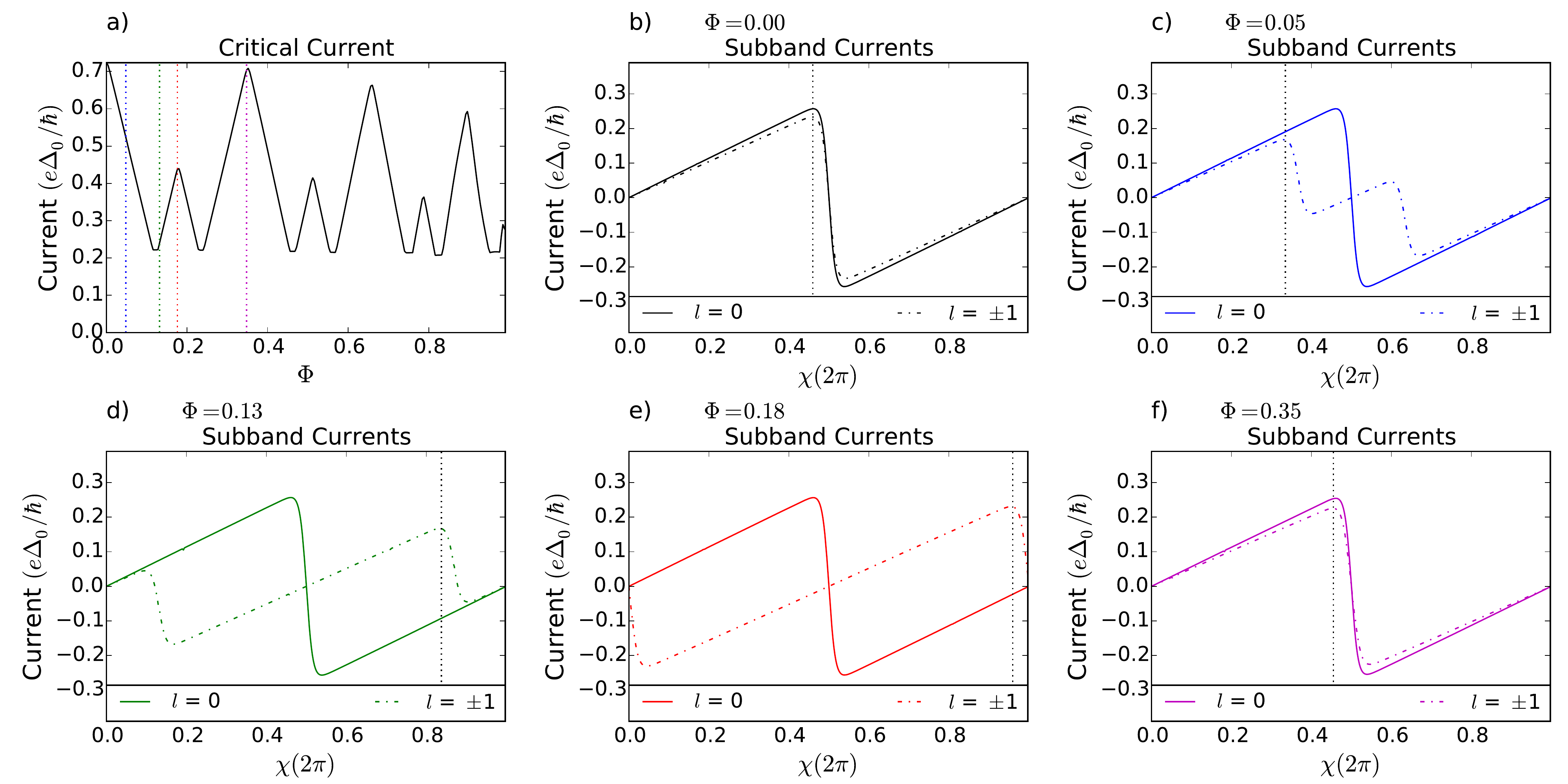}
	\end{center}
	\caption{a) Critical current $I_c$ vs normalized axial magnetic flux $\Phi$ of a 500 nm junction with $\mu = 8.5 \; \mathrm{meV}$. The subbands with $l=-1,0,1$ are occupied and contribute to $I_c$. Oscillation of $I_c$ with $\Phi$ is observed, which is not periodic in $\Phi$: the positions of the peaks get closer together as $\Phi$ is increased. At flux points indicated with vertical dotted lines, individual subband currents are plotted in panels b-f vs the superconducting phase $\chi$. b-f) CPR for $l=0$ (solid lines) and $l = \pm 1$ (dash-dotted lines) subbands. The current due to the $l=1$ subband is equal to that of the $l=-1$ subband for all $\chi$ and $\Phi$ values. Here, the contribution due to only one of the two subbands is shown for clarity. The vertical dotted lines indicate the phase $\chi$ at which the critical current occurs. Note the difference in the y-axis scale between panel a and the other panels. The following parameters were used in all plots: $L = 500\; \mathrm{nm}, \mu = 8.5 \; \mathrm{meV}, R = 30 \; \mathrm{nm}, T = 100\; \mathrm{mK}$.}
	\label{fig:fig5_slices_8p5}
\end{figure*}
\indent Since the junction length is greater than the healing length of the populated subbands ($L = 500 \; \mathrm{nm} \gtrsim \xi^0_{n,l}$), the long junction limit applies. Many bound states are present in the junction ($\sim 12$). In figure \ref{fig:fig4_currents}a we show the bound state current $I_{n,l}$, continuum current $J_{n,l}$, and their sum, for the subband $(n,l) = (0,1)$ at zero magnetic field. As expected of a long junction \cite{bardeen_long_j, bagwell}, $I_{n,l}$ and $J_{n,l}$ are of the same order of magnitude, and the CPR is triangular in shape. An additional group of 4 bound states ($\Psi^+, \Psi^-, \overbar {\Psi}^+, \overbar {\Psi}^-$) appear in the junction at $\chi = 0.05$, and exit at $\chi = 0.95$, giving rise to discontinuities in $I_{n,l}$, $J_{n,l}$. However, the total subband current $I_{n,l} + J_{n,l}$ is always a smooth function of $\chi$. It is maximal near $\chi = \pi$, (exactly at $\chi = \pi$ at zero temperature) regardless of the junction length \cite{bagwell}. Note that the continuum current is zero at $\chi = \pi$.\\ 
\indent In figure \ref{fig:fig4_currents}b we show the subband current $I_{n,l} + J_{n,l}$ as a function of the magnetic flux. As the flux is increased from zero, two discontinuities develop in the subband current (shown for $\Phi = 0.10$). The bound state current is modified as the eigenenergies corresponding to states $(\Psi^+, \overbar{\Psi}^+)$ are shifted in phase in the opposite direction to those of states $(\Psi^-, \overbar{\Psi}^-)$, similarly to figure \ref{fig:fig3_BS_phase_shift}b. An equivalent process happens for the continuum current, as explained in Eq. \ref{eq:D_cont_approx}. As a result, the subband current $I_{n,l} + J_{n,l}$ also shows two discontinuities, and is no longer necessarily maximal near $\chi = \pi$.\\
\indent The amounts of the phase shifts in $I_{n,l}$ and $J_{n,l}$ depend on the quantity $(k^e_{n,l} - k^h_{n,l})L$. The wavenumbers $k^e_{n,l}, k^h_{n,l}$ are subband parameters defined in Eq. \ref{eq:wavenumbers}, and the length $L$ is device dependent. Therefore, the fluxes at which phase shifts equal integer multiples of $2\pi$ need not occur at integer multiples of $\Phi_0 = (h/e)$ or $\Phi_0/2$; they can occur at any value of $\Phi$. An example is shown in figure \ref{fig:fig4_currents}b for $\Phi = 0.35$, where the phase shifts equal $2\pi$ and the CPR recovers its shape at $\Phi = 0$. Notice, however, that the maximal value of the subband current at $\Phi = 0.35$ is smaller than at $\Phi = 0$. This can be intuitively understood as follows: as the flux increases, the effective wavenumbers of the electrons and holes change according to Eq. \ref{eq:wavenumbers}. It can be seen that the average momentum of the electron-hole pair and therefore the healing length $\xi^0_{n,l}$ are always smaller for higher fluxes. Since the magnitude of Josephson current in a long junction scales approximately linearly \cite{Furusaki1991} with $\xi^0_{n,l}/ L$, it is suppressed at higher fluxes. This suppression is stronger near the depopulation point of a given subband, where the average momentum decreases significantly.\\
\subsection{Interference due to a few subbands}
\label{sec:32_few}
In order to elucidate the mechanism of the Josephson interference between subbands, we show in figure \ref{fig:fig5_slices_8p5}a the critical current versus axial flux of the junction studied in section \ref{sec:31_one_subband}. The length $L = 500\; \mathrm{nm}$ is chosen because it allows for a relatively large amount of phase pickup, since the phase pickup is proportional to the length of the junction ($(k^e_{n,l} - k^h_{n,l})L$ in Eq. \ref{eq:quantization-rule}). Hence, several oscillations of the critical current occur prior to the depopulation of the $|l|=1$ subbands at $\Phi = 1$.
\begin{figure*}[t!]
	\begin{center}
		\includegraphics[width=7in]{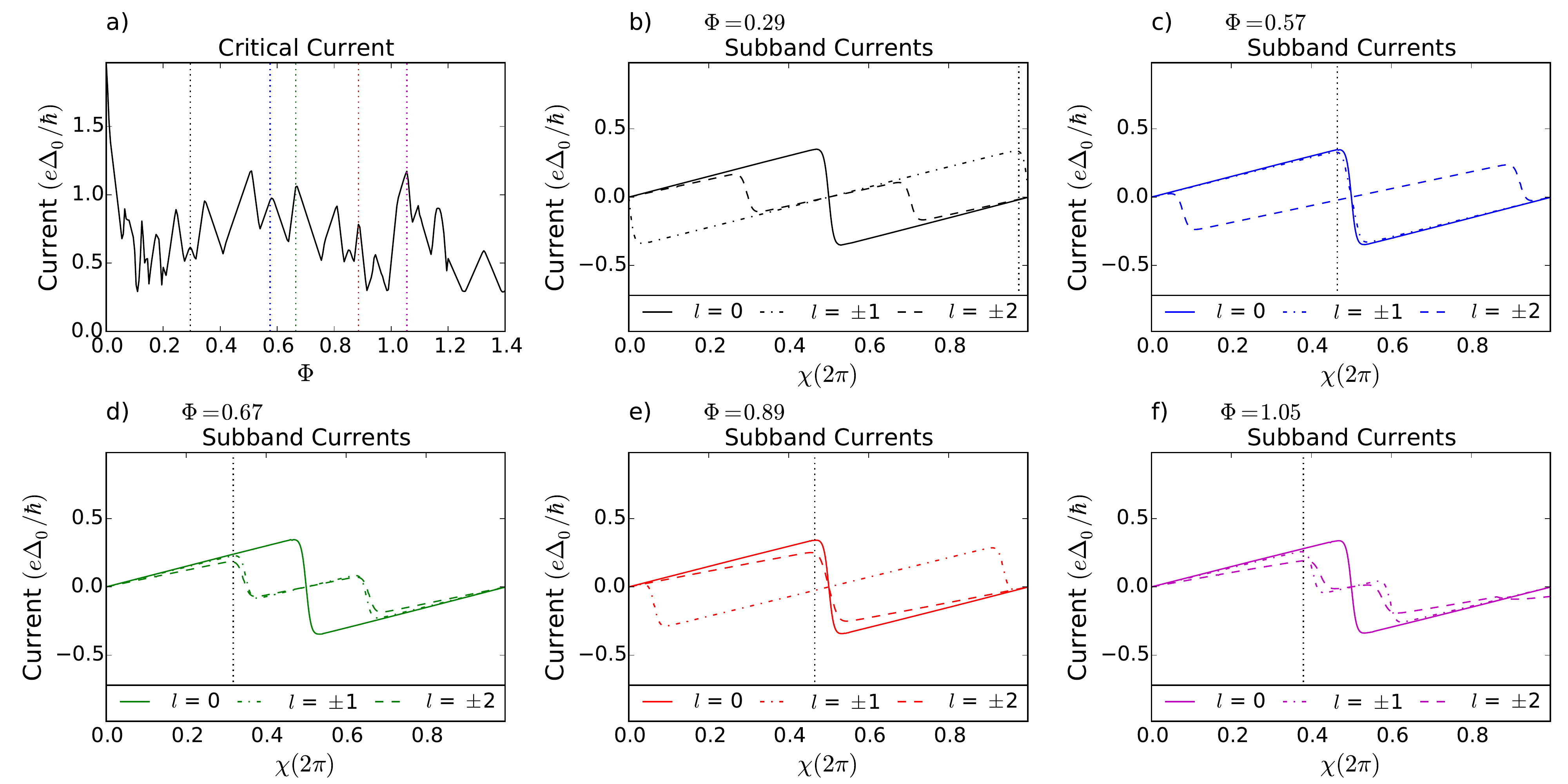}
	\end{center}
	\caption{a) Critical current $I_c$ vs normalized axial magnetic flux $\Phi$ of a 500 nm junction with $\mu = 20 \; \mathrm{meV}$. The subbands with $|l| \leq 3$ are occupied at zero flux. The subbands with $|l| = 3,2$ depopulate at $\Phi = 0.2, 1.2$, respectively. An aperiodic oscillation of $I_c$ vs $\Phi$ is observed. At flux points indicated with vertical dotted lines, individual subband currents are plotted in panels b-f vs the superconducting phase $\chi$. b-f) CPR for individual subbands, displaying different configurations which can lead to a peak in $I_c$ vs $\Phi$. The vertical dotted lines indicate the phase $\chi$ at which the critical current occurs. Note the difference in the y-axis scale between panel a and the other panels. The following parameters were used in all panels: $L = 500\; \mathrm{nm}, \mu = 20 \; \mathrm{meV}, R = 30 \; \mathrm{nm}, T = 100\; \mathrm{mK}$.}
	\label{fig:fig6_slices_20}
\end{figure*}
\indent The supercurrent of each subband is shown versus the phase difference $\chi$ in figure \ref{fig:fig5_slices_8p5} panels b-f, at particular values of the magnetic flux. The $l=1$ subband current equals that of $l=-1$ at all fluxes. At zero flux the current of each subband is maximal near $\chi = \pi$ (exactly $\chi = \pi$ for zero temperature), as discussed in section \ref{sec:31_one_subband}. This can be clearly seen in panel b. The total current of the junction is the sum of the contributions from the $l = -1,0,1$ subbands, and is therefore roughly three times the contribution of each. The dotted vertical lines show the phase at which the critical current occurs.\\
\indent As the flux is increased, the CRP of the $|l| = 1$ subbands are modified, similarly to figure \ref{fig:fig4_currents}a. The critical current of the junction decreases, since the $|l| = 0,1$ subbands no longer interfere constructively (panel c). At $\Phi = 0.13$ the maximal current switches from a phase $\chi < \pi$ to $\chi > \pi$, as shown in panel d. This is a feature of the triangular CPR. The critical current increases until $\Phi = 0.18$ (panel e), at which point the junction current is maximal near $\chi = 2 \pi$. We call this a peak a secondary peak as it occurs roughly in the middle of the main period of oscillations (see discussion below on periodicity), when the magnetic phase pickup of the $|l|= 1$ subbands equals roughly $\pi$. The magnitude of this peak is roughly two thirds the total current as zero field, as the $l = \pm 1$ subbands contribute maximally, and the $l=0$ subband current is close to zero. The process reverses itself for $\Phi > 0.18$, until at $\Phi = 0.35$ the phase pickup of the $|l|= 1$ subbands equals $2\pi$ and all subbands interfere constructively again (panel f). We refer to the peak at $\Phi = 0.35$ a primary peak. Other primary peaks occur at $\Phi = 0.66, 0.89$. As in figure \ref{fig:fig4_currents}, the contribution of the $|l|= 1$ subbands decrease as $\Phi$ is increased, because the decrease in the average quasiparticle momentum. This is the mechanism behind the slow decay of the magnitude of the primary peaks as flux increases.\\
\indent \textit{Aperiodicity --} The $|l| =0, 1$ subbands interfere constructively when the magnetic phase pickup of the $|l|= 1$ subbands equals an integer multiple of $2\pi$. This corresponds to the main period of the critical current oscillations with $\Phi$, which we estimate below. In other words, we want to find $\Phi$ such that
\begin{equation}
\left. (k^e_{n,l} - k^h_{n,l})L \right|_{\Phi} - \left.(k^e_{n,l} - k^h_{n,l})L \right|_{\Phi =0} = 2j\pi,
\end{equation}
for integer $j$. The wavenumbers $k^e_{n,l}, k^h_{n,l}$ are defined in Eq. \ref{eq:wavenumbers}. For the general case this is not easy to do analytically, as $k^e_{n,l}, k^h_{n,l}$ themselves depend on the flux through the effective chemical potentials $\mu^{e,h}_{n,l}$. However, we can get an estimate of the expected period by invoking the Andreev approximation (Eq. \ref{eq:wavenum-approx}), and assuming the effective Fermi velocities do not depend on the flux, so can be evaluated at some fixed $\Phi$, e.g. $\Phi = 0$. These are reasonable assumptions when the flux is much smaller than the depopulation point of a given subband (e.g., $\Phi = 1$ for $|l| = 1$ subbands in the example of figure \ref{fig:fig5_slices_8p5}). The result for the position of the first primary peak $\Phi_1$ is:
\begin{equation}
\Phi_1 = \pi v_{n,l} m^* R^2 / (\hbar l L),
\label{eq:period_andreev}
\end{equation}
where $v_{n,l}$ is defined below Eq. \ref{eq:wavenum-approx}. For the example of figure \ref{fig:fig5_slices_8p5}, this evaluates to $\Phi_1 = 0.36$, deviating from the numerically calculated value by only $3\%$. However, we see that the numerical positions of the next primary peaks at $0.66$, $0.89$ cannot be accurately approximated as integer multiples of $\Phi_1$: the period becomes shorter as $\Phi$ is increased. This is because the field dependence of $v_{n,l}$ cannot be ignored at higher values of $\Phi$, and the Andreev approximation breaks down. Intuitively, the effective Fermi velocity is noticeably lower at higher fields, resulting in more time spent in the junction by the Andreev pair and more phase pickup, therefore a smaller period for $I_c$ oscillations. Consequently, even in the simple case of a few subbands, the oscillations of $I_c$ versus $\Phi$ are not strictly periodic.\\
\subsection{Interference due to many subbands}
\label{sec:33_many}
A higher chemical potential results in the occupation of a greater number of angular momentum subbands. The rich interplay between the different $l$-subbands results in a complex pattern of the oscillation of $I_c$ with $\Phi$. Assume subbands with $|l|$ up to $\tilde{l}$ are occupied. Each $|l|$ subband's supercurrent oscillates with a flux dependent `period' approximated by Eq. \ref{eq:period_andreev}, which depends on the subband velocity $v_{n,l}$, and is therefore generally anharmonic with other subbands. Typically, a peak in $I_c$ as a function of $\Phi$ occurs under one of two circumstances: (i) when some (at least two) of the subbands with different $|l|$ values interfere constructively, or (ii) when the critical current occurs near $\chi = 2\pi$, a secondary peak is formed as described in figure \ref{fig:fig5_slices_8p5}. There are $\left(
\begin{array}{c}
\tilde{l} \\
 2
\end{array}
\right)$ choices for pairs of constructively interfering subbands (parentheses indicate the binomial coefficient). As each $|l|$ subband's oscillations can be anharmonic with those of all other subbands, it follows that there are $\left(
\begin{array}{c}
\tilde{l} \\
 2
\end{array}
\right)$ different flux-dependant `periods' in the $I_c$ oscillations due to condition (i), with another $\tilde{l}$ due to condition (ii). The $I_c$ curves therefore can display complex, aperiodic structures.\\
\indent In figure \ref{fig:fig6_slices_20}a we plot an example of an $I_c$ versus $\Phi$ curve for a junction with the same parameters as that of figure \ref{fig:fig5_slices_8p5}, except the chemical potential is raised from 8.5 meV to 20 meV. At zero flux, subbands up to $|l| = 3$ are occupied. The $|l|=3$ states depopulate at $\Phi = 0.2$, and the $|l|=2$ states at $\Phi = 1.2$. As an example of configurations that can give rise to a peak in $I_c$, subband CPRs are shown in figure \ref{fig:fig6_slices_20} panels b-f, for 5 peaks indicated in panel a with vertical dotted lines. The peak at $\Phi = 0.29$ (panel b) satisfies condition (ii), while the other examples are due to constructive interference of two subbands, i.e. condition (i): subbands with $|l| = 0,1$ at $\Phi = 0.57$ (panel c),  $|l| = 0,2$ at $\Phi = 0.89$ (panel e), and $|l| = 1,2$ at $\Phi = 0.67$ and $\Phi = 1.05$ (panels d,f).
\begin{figure}[t]
	\begin{center}
		\includegraphics[width=3.5in]{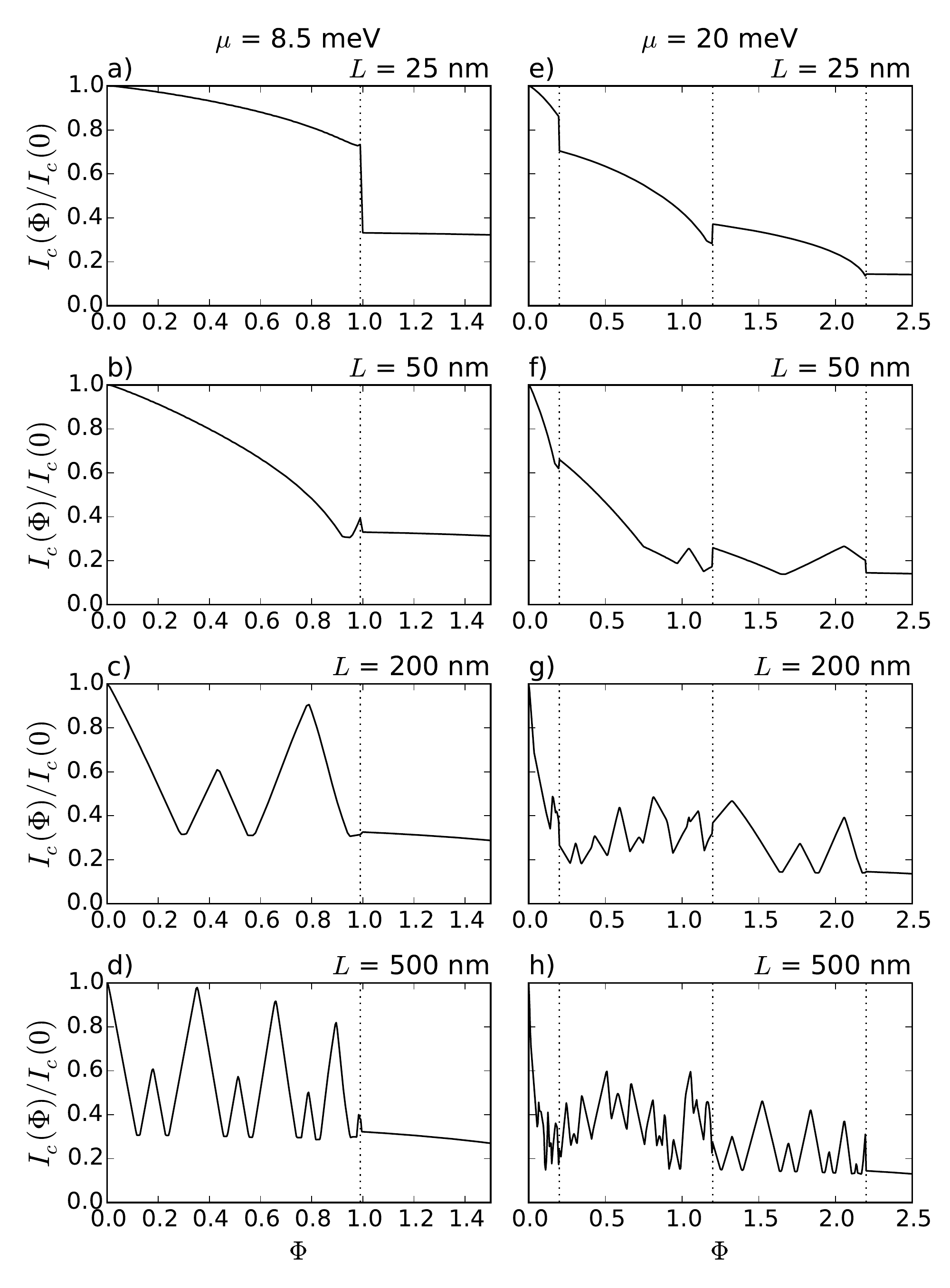}
	\end{center}
	\caption{Normalized critical current of the junction versus the normalized magnetic flux $\Phi = (\pi B_\parallel R^2)/(h/e)$, for $\mu = 8.5\; \mathrm{meV}$ (panels a-d) and $\mu = 20\; \mathrm{meV}$ (panels e-h), and several values of the junction length $L$. The following parameters were used: $R = 30\; \mathrm{nm}, T = 100\; \mathrm{mK}$. In panels a-d, the vertical line at $\Phi = 1.0$ indicates the depopulation of the $|l| = 1$ subbands. Similarly in panels e-h the vertical lines at $\Phi = 0.2, 1.2, 2.2$ indicate the depopulation of the $|l| = 3,2,1$ subbands, respectively.}
	\label{fig:fig7_res}
\end{figure}
\subsection{Effect of junction length}
\label{sec:effect_of_length}
\indent We now discuss the effect of the junction length $L$ on the pattern of $I_c$ oscillations. The left column in figure \ref{fig:fig7_res} (panels a-d) shows the numerically obtained $I_c$ versus $\Phi$ for a junction with $\mu = 8.5\; \mathrm{meV}$, as $L$ is varied. All other junction parameters are the same as in figure \ref{fig:fig5_slices_8p5}. The critical current is normalized to its value at zero magnetic flux. Two processes affect the behaviour of $I_c$: (i) the depopulation of the $|l| = 1$ subbands at $\Phi = 1$, shown by vertical dotted lines in panels a-d, and (ii) the Josephson interference effect described above. The first of these processes results in step-like discontinuities in $I_c$ at $\Phi = 1.0$, a drop to roughly one-third of the zero-field $I_c$ value as the $l = -1, 1$ subbands depopulate. Since the $l = 0$ subband does not couple to the axial flux, $I_c$ is almost constant above $\Phi = 1$. The slow decay of the primary peak heights below $\Phi = 1$, and of $I_c$ above $\Phi = 1$, are both due to the decreasing average momentum at higher fluxes, as discussed in section \ref{sec:31_one_subband}. For a short junction with $L = 25\; \mathrm{nm}$ (panel a) and for $\Phi < 1.0$, the phase shifts in the CPR of the $|l| = 1$ subband are small (the phase shifts are dependent on $(k^e_{n,l} - k^h_{n,l})L$), but become more significant close to $\Phi = 1$, where the axial velocities of the quasiparticles are smaller and the time of flight across the junction longer. This results in the observed decrease in $I_c$ before the step-like discontinuity. As  $L$ is increased to 50 nm, the value of this phase shift increases and modulation due to interference starts to emerge. For $L = 200\; \mathrm{nm}$ (figure \ref{fig:fig7_res} panel c), the first primary peak occurs at $\Phi = 0.79$, and for $L = 500\; \mathrm{nm}$ (panel d) at $\Phi = 0.35$. The decrease by a factor of $2.26$ in the position of the first primary peak is approximately equal (within $11 \%$) to the reciprocal ratio of lengths $L$, as expected from the estimate in Eq. \ref{eq:period_andreev}.\\
\indent The right column of figure \ref{fig:fig7_res} (panels e-h) shows the $I_c$ versus $\Phi$ curves for a junction with $\mu = 20 \; \mathrm{meV}$. The subbands with $|l| = 3,2,1$ depopulate at $\Phi = 0.2,1.2,2.2 $ respectively (vertical dotted lines in panels e-h), resulting in step-like discontinuities in $I_c$ at those flux values. Similarly to above, the behaviour of $I_c$ is dominated by this effect for a $25\; \mathrm{nm}$ junction (panel e), but as $L$ is increased the Josephson interference becomes visible (panels f-h). Note that between the flux values $\Phi = 1.2$ and $\Phi = 2.2$ the $I_c$ curves in the right column of figure \ref{fig:fig7_res} are qualitatively similar to their counterparts in the left column, because similarly to the $\mu = 8.5\; \mathrm{meV}$ case, subbands with $|l| \leq 1$ are occupied within this flux window. Note that for $\mu = 20\; \mathrm{meV}$, $I_c(0)$ is much larger than for the $\mu = 8.5\; \mathrm{meV}$ case; the oscillation amplitude appears to be smaller in panels (e-h) only because the relative contribution of each subband to the total current is smaller when there are more subbands occupied. \\
\indent In summary, for a short junction, the depopulation of subbands is more visible than the Josephson interference effect. However, as the junction length is increased, the interference effect becomes apparent. The periods of oscillation decrease slightly as flux is increased. For a junction with a low chemical potential (only a few transverse subbands occupied, e.g. an $SNS$ point-contact \cite{pointcont1,Furusaki1999,bagwell_one_dim_2}), the pattern of $I_c$ modulation is simpler and the period longer than the case of a high chemical potential (many transverse subbands occupied). A long, low-$\mu$ junction is optimal for experimental observation of this Josephson interference effect.
\section{Discussion}
\label{sec:4_disc}
We have described the theory of a previously unstudied form of the Josephson interference effect that can occur in nanoscale $SNS$ junctions due to the coupling of the orbital angular momentum of transverse electronic subbands with an axial magnetic flux. We found in section \ref{sec:effect_of_length} the regimes in which this interference effect dominates the $I_c$ vs $\Phi$ characteristics of the junction. An idealized model of an $SNS$ junction was used, with several simplifying assumptions, in order to elucidate the mechanism of the effect. We discuss generalizations of the model below, in particular those modifications that may be necessary to directly model experimental devices.\\
\indent \textit{FWVM and barriers at the interfaces} -- No barriers were assumed at the $S$-$N$ interfaces, and FWVM was neglected. The effective mass for electrons $m^*$ was assumed to be uniform throughout the junction. These assumptions allowed Kulik's method of matching the wavefunctions at the interfaces to be used to calculate the bound state and the continuum currents. We stress that the basic mechanism of the orbital interference effect, i.e. the modification of the $N$-section wavenumbers in the presence of the axial field, is independent of FWVM and interfacial barriers. Therefore, the main features of the $I_c$ oscillations (periodicity, amplitude) should only be modified by FWVM and barriers as higher order corrections. The exact shape of the junction CPR and the $I_c$ vs $\Phi$ curves, however, depends on the details of the interfaces. Accurate modeling of experiments must take this into account, based on the material and interfacial properties specific to a particular experimental implementation. \\
\indent As was discovered in studies of Andreev reflection at Nb-InAs interfaces \cite{chrestin_sm_oscillations,schapers_book_2}, FWVM modifies the CPR of Superconductor/Semiconductor/Superconductor junctions. In the general case, where FWVM and barriers are included at the $S$-$N$ interfaces, the bound state energies and the continuum current must be calculated from the transmission matrix formalism \cite{tang}, in which the transmission matrix includes a normal (specular) reflection coefficient as well as an Andreev (retro) reflection coefficient. The values of these coefficients depend on the material details of the junction. The CPRs of junctions with FWVM \cite{tang} and barriers \cite{Furusaki1999,heida_phd,bagwell} have been previously calculated (numerically) at zero magnetic field. The effect of both mechanisms is to make the CPR more closely resemble a sinusoidal curve. Since the $N$-section wavenumbers have the same coupling to the axial field shown in Eq. \ref{eq:wavenumbers} regardless of FWVM or barriers, we expect the phase-shift mechanism leading to interference to remain. However, the shape of the oscillations in $I_c$ should appear more sinusoidal, following the shape of the CPR.\\
 \indent Normal reflection of the quasiparticle wavefunctions from the $S$-$N$ barriers will result in a higher order correction to the periodicity of the interference effect. This is due to an increased average phase pickup as the quasiparticle spends more time in the junction. On the other hand, this should also lead to some randomization of the phase. The former would lead to shorter period oscillations (i.e the effect occurs at lower fields), while the latter will partially smear out the interference effect, reducing the amplitude of oscillations. Similar effects are expected from elastic backscattering occurring in the $N$-section in non-ballistic junctions. These considerations are beyond the scope of this paper, and are left to future work.\\
\indent \textit{General form of $\Delta$} -- A spatially uniform pairing potential was assumed in the $S$-sections at all magnetic fields (Eq. \ref{eq:delta}). This was justified by assuming cylindrical $S$-sections, and restricting the cylinder diameter to be smaller than the superconducting coherence length in the $S$-sections. However, experimental fabrication of nanoscale $SNS$ junctions is usually done by evaporating or sputtering metallic (e.g. Al or Nb) thin film contacts onto a semiconducting nanowire. In this case, the geometry of the $S$-section is not cylindrical but $\Omega$ shaped. The lack of cylindrical symmetry necessitates, in principle, a 3-dimensional numerical calculation of $\Delta(\bo r)$ using self-consistent methods. However, as long as the $N$-section can be assumed to be cylindrically symmetric, our model should closely approximate the experimental situation. This is because the interference effect depends mainly on the eigensolutions in the $N$-section, particularly the orbital angular momentum states and their coupling to the flux. The details of the eigensolutions in the $S$-section do not play a direct role, other than asserting the form of the wavefunction ansatz (Eq. \ref{eq:psi_plus}) is valid. Similarly, the axial field can induce a non-uniformity in $\Delta$. If the thickness of the metallic film becomes larger than the $S$-section coherence length, fluxoid quantization can result in a $\theta$-dependent phase for $\Delta$ in the presence of the field. Inserting a $\theta$-dependent $\Delta$ in the BdG equations (Eq. \ref{eq:bdg}) will affect the bound state solutions, likely requiring a 3-dimensional numerical solution. However, we still expect the interference effect to depend mainly on the states in the $N$-section.  \\
\indent \textit{General radial wavefunctions} -- The shell-conduction model was used in order to simplify computations, and to help gain intuitive insight into the problem; it is not strictly necessary for the main arguments of the paper. Indeed, we found in section~\ref{sec:semiclassical} that the semiclassical phase shift $\delta_{sc}$ is only weakly dependent on the radius $R$, so the interference effect should be present for general radial wavefunctions. In future work, the radial wavefunctions in the $N$-section, $\phi_{n,l}$, and the corresponding single particle energies in the presence of the field will be numerically calculated, yielding the appropriate wavenumbers $k^e_{n,l}, k^h_{n,l}$ for electron- and hole-like solutions. We expect the term $(k^e_{n,l} - k^h_{n,l})L$ appearing in Eqs. \ref{eq:quantization-rule}, \ref{eq:continuum-current} will continue to result phase shifts similar to those seen in the present model. \\  
\indent \textit{Zeeman and Spin-orbit effects} --  In order to study the the orbital Josephson interference effect in isolation, Zeeman and spin-orbit effects were neglected in our analysis. It is useful to ask under what circumstances should the orbital effects or the Zeeman + spin-orbit effects dominate? The critical current of a short, InSb $SNS$ junction, including spin-orbit and Zeeman effects, was studied in Ref.\cite{yokoyama_spin_orbit_zeeman}. The bound state energies were solved, and a phase-shift was observed in the energy-versus-phase curves due to the Zeeman effect. Similar to the mechanism described in our analysis, the Zeeman effect modifies the $N$-section wavenumbers, but based on the spin state rather than the orbital state. This results in an oscillation of $I_c$, with the first minimum occurring at $B_{min} = \hbar v_F/ (g \mu_B L)$, where $v_F$ is the Fermi velocity, $g$ is the effective Land{\'e} g-factor, and $\mu_B$ is the Bohr magneton. For a 200 nm InSb junction with $|g| = 50$, this evaluates to $B_{min} = 0.5$ T for $v_F$ corresponding to $\mu = 10$ meV, but for other materials $B_{min}$ is typically much larger. Considering InAs with a moderately large $|g| = 10$ gives  $B_{min} = 2.1$ T (again for $L=200$ nm and $\mu = 10$ meV). Since we used the effective mass for InAs in our calculations, we can directly compare with the results of figure \ref{fig:fig7_res}c for a 200 nm long junction. The orbital effect should dominate in this case, as the first minimum of $I_c$ is at a flux corresponding to $B_\parallel \simeq 0.4$ T. The consequence of the inclusion of the spin-orbit coupling is a smaller correction: the so called anomalous Josephson effect, in which the current is no longer an even function of the superconducting phase: $I(\chi) \neq I(-\chi)$.\\
\indent We conclude that for InSb devices, the Zeeman effect could easily dominate. This is especially true if either of the following conditions hold: (i) If only $l= 0$ subbands are occupied (i.e. small chemical potential), or (ii) in a perpendicular field, where the Zeeman effect is present but the orbital effect is suppressed. On the other hand, for most low spin-orbit semiconductor materials, $g \sim 2$ and we would expect the orbital subband effect to dominate in an axial field experiment, unless only $l = 0$ subbands are occupied. \\
\indent \textit{Magnetic depairing} -- Field-induced depairing suppresses both superconductivity in the $S$-sections and proximity superconductivity in the $N$-section \cite{proximityEffectInN, proximityEffectInAs}. For a type-II superconductor with a relatively large gap such as Nb, we can assume depairing in the $N$-section should dominate. For diffusive junctions in the narrow junction limit, the Usadel equations predict a monotonic Gaussian decay $I_c \propto \text{exp}(-\alpha \Phi^2 / \Phi_0^2)$ for a perpendicular magnetic flux $\Phi$ through the $N$-section \cite{Hammer_non-ideal-interface}, where $\alpha \approx 0.24$ is a numerical constant. A similar effect should apply to the axial field case, except with a slower magnetic field decay due to a smaller cross-sectional area (smaller flux). This is expected to produce a monotonic decay envelope superimposed on the critical current oscillations, and should be taken into account when modeling experimental data. \\
\indent \textit{Summary} -- The idealized model studied here serves to demonstrate a novel form of Josephson interference due to orbital angular momentum states, with the unusual property of flux-aperiodic oscillations. Extensions to the model discussed above, most importantly FWVM and barriers at the $S$-$N$ interfaces, will be useful for describing experimental implementations. 

\begin{acknowledgements}
We thank A. J. Leggett and S. Frolov for helpful discussions. This work was supported by NSERC and the Ontario Ministry for Research and Innovation.
\end{acknowledgements}

\appendix
\section{Transmission formalism for the continuum current}
\label{sec:appendix_cont}
The electrical current transmission amplitudes for the continuum states are obtained by matching the solutions of the BdG equation (Eq. \ref{eq:bdg}) in the three regions of the cylinder ($x<-L/2,\, |x|<L/2,\, x>L/2$) while assuming an incident ``source term" on the $S$-$N$ interface at $x = -L/2$ with energy $|E| > \Delta_0$. These transmission amplitudes are then used to calculate the continuum current. This section follows appendices A,B in Ref. \cite{bagwell}, but is generalized to account for finite magnetic field and orbital angular momenta.\\
\indent Hereafter all quantities are presumed to pertain to one subband, $(n,l)$, unless explicitly stated otherwise. We drop the subscripts $n,l$ for simplicity. Consider the quasiparticle excitation spectrum of the BdG Hamiltonian. In the left $S$-section ($x < -L/2$), the electron-like solutions with energy $E>\Delta_0>0$ are given by
\begin{equation}
\Psi^e = \left( \begin{array}{c} u(x,\rho,\theta) \\ v(x,\rho,\theta) \end{array} \right) = \left(\begin{array}{c} u_0 e^{i\chi_L} \\ v_0 \end{array}\right) \psi^{L,e} (x,\rho) e^{il\theta},
\label{eq:app_psi_e}
\end{equation}
and the hole-like solutions by
\begin{equation}
\Psi^h = \left( \begin{array}{c} u(x,\rho,\theta) \\ v(x,\rho,\theta) \end{array} \right) = \left(\begin{array}{c} v_0 e^{i\chi_L} \\ u_0 \end{array}\right) \psi^{L,h} (x,\rho) e^{il\theta}.
\end{equation}
Here, $u_0, v_0$ are given in Eq. \ref{eq:coherence-factors}, and $\psi^{L,e}, \psi^{L,h}$ have the form given above Eq. \ref{eq:psi_plus}, with different expansion coefficients $\beta_p$ for the electron- and hole-like wavefunctions. We do not reproduce these coefficients here as they do not enter the calculation. Let the source term $\Psi^e$ be incident from the left on the $S$-$N$ interface at $x = -L/2$. The wavefunctions generated due to this source term are grouped into two categories (depending on which $S$-section they belong to), with coefficients $B,C$, following Bagwell's notation \cite{bagwell}. The first group pertains to the left $S$-section. The electron-like source term incident on the $S$-$N$ interface is
\begin{equation}
\left(\begin{array}{c} u_0 e^{i\chi_L} \\ v_0 \end{array}\right) \psi^{L,e} (x+\frac{L}{2},\rho) e^{il\theta}\quad (x < -\frac{L}{2}).
\label{eq:app_psi_e_incident}
\end{equation}
The Andreev reflected hole-like wavefunction is
\begin{equation}
\left(B - \frac{v_0}{u_0}\right) \left( \begin{array}{c} v_0 e^{i\chi_L} \\ u_0 \end{array} \right) \psi^{L,h} (x+\frac{L}{2},\rho) e^{il\theta}\quad (x < -\frac{L}{2}).
\label{eq:app_andreev_wf}
\end{equation}
For the electrons in the normal region, the wavefunction is
\begin{equation}
\begin{array}{ll}
\left(B - \frac{v_0}{u_0} + \frac{u_0}{v_0}\right) \left( \begin{array}{c} v_0 e^{i\chi_L} \\ 0 \end{array} \right) e^{i(k^e(E))\times(x+\frac{L}{2})} & \phi(\rho) e^{il\theta}\\
\; & (|x| < \frac{L}{2}), \end{array}
\label{eq:app_elec_left}
\end{equation}
and for the holes
\begin{equation}
B \left( \begin{array}{c} 0 \\ u_0 \end{array} \right) e^{i(k^h(E))\times(x+\frac{L}{2})} \phi(\rho) e^{il\theta}\quad (|x| < \frac{L}{2}).
\label{eq:app_hole_left}
\end{equation}
Here, $\phi$ is radial part of the wavefunction in the $N$-section, defined below Eq. \ref{eq:bdg}. The explicit energy dependence of the wavenumbers $k^e, k^h$ is given in Eq. \ref{eq:wavenumbers}.\\
\indent The transmitted wavefunction into the right contact is
\begin{equation}
C  \left( \begin{array}{c} u_0 e^{i\chi_R} \\ v_0 \end{array} \right) \psi^{L,e} (x-\frac{L}{2},\rho) e^{il\theta}\quad (x > \frac{L}{2}).
\end{equation}
This is supported by electrons in the $N$-section with the wavefunction
\begin{equation}
C \left( \begin{array}{c} u_0 e^{i\chi_R} \\ 0 \end{array} \right) e^{i(k^e(E))\times(x-\frac{L}{2})} \phi(\rho) e^{il\theta}\quad (|x| < \frac{L}{2}),
\label{eq:app_elec_right}
\end{equation}
and holes
\begin{equation}
C \left( \begin{array}{c} 0 \\ v_0 \end{array} \right) e^{i(k^h(E))\times(x-\frac{L}{2})} \phi(\rho) e^{il\theta}\quad (|x| < \frac{L}{2}).
\label{eq:app_hole_right}
\end{equation}
The normal reflection processes have not been considered, as no FWVM or barriers are assumed at the $S$-$N$ interfaces.\\
\indent The coefficients $B,C$ can be found by connecting together Eqs. \ref{eq:app_elec_left}, \ref{eq:app_elec_right}, and Eqs. \ref{eq:app_hole_left}, \ref{eq:app_hole_right} at any point $x = a$ within the $N$-section or at the $S$-$N$ interfaces. We use $a = -L/2$. The result is
\begin{equation}
C = \frac{1 - \frac{v_0^2}{u_0^2}}{e^{i\chi}e^{-ik^e(E)L} - \frac{v_0^2}{u_0^2}e^{-ik^h(E)L}},
\end{equation}
where $\chi = \chi_R - \chi_L$. By definition, $C$ is the transmission amplitude from the left contact to the right contact due to an electron-like incident source term with energy $E>0$. The corresponding transmission coefficient $T^e_{L\rightarrow R}(E,\chi)$ is
\begin{equation}
T^e_{L \rightarrow R} (E,\chi) = |C|^2 = \frac{|u_0^2 - v_0^2|^2}{F^+(E,-\chi)},
\label{eq:app_t_lr}
\end{equation}
with the function $F^+$ given below Eq. \ref{eq:coherence-factors}.\\
\indent The coefficient for transmission from right to left (due to a left-moving source) can be found by making the transformation $\chi \rightarrow -\chi$ in the above formula. That is,
\begin{equation}
T^e_{R \rightarrow L} (E,\chi) = \frac{|u_0^2 - v_0^2|^2}{F^+(E,\chi)}.
\label{eq:app_t_rl}
\end{equation}
Repeating these calculations for an electron-like source term with energy $E<-\Delta_0<0$ shows that Eqs. \ref{eq:app_t_lr}, \ref{eq:app_t_rl} give the correct transmission coefficients for the negative energy case as well.\\
\indent Similar to the case of the bound states, if $(u,v)^T$ is a solution at energy $E$, then $(-v^*, u^*)^T$ gives a solution at energy $-E$. Both types of solutions must be taken into account when calculating the total transmission coefficients. Consider the (left-moving) source term $\overbar {\Psi^e}$ obtained by applying the above transformation on Eq. \ref{eq:app_psi_e}:
\begin{equation}
\overbar {\Psi^e} = \left( \begin{array}{c} -v_0 \\ u_0 e^{-i\chi_L} \end{array} \right) \left( \psi^{L,e}(x,\rho)\right)^*e^{-il\theta}.
\end{equation}
All relevant wavefunctions due to this source term (i.e. the Andreev reflected, $N$-section electron- and hole-like, and transmitted wavefunctions) can be constructed by applying the  transformation $(u,v)^T \rightarrow (-v^*, u^*)^T$ to Eqs. \ref{eq:app_andreev_wf}\---\ref{eq:app_hole_right}. Crucially, the resulting wavefunctions contain the wavenumbers $k^e(E), k^h(E)$, while having energy $-E$. Repeating the above calculation, the transmission coefficient $ \overbar{ T^e_{ R \rightarrow L}}$ due to the source term $\overbar {\Psi^e}$ is found:
\begin{equation}
\overbar{ T^e_{ R \rightarrow L}}(-E,\chi) = \frac{|u_0^2 - v_0^2|^2}{F^+(E,-\chi)},
\end{equation}
or equivalently,
\begin{align}
\overbar{ T^e_{ L \rightarrow R}}(E,\chi) = &\frac{|u_0^2 - v_0^2|^2}{F^-(E,\chi)},\\
\overbar{ T^e_{ R \rightarrow L}}(E,\chi) = &\frac{|u_0^2 - v_0^2|^2}{F^-(E,-\chi)},
\end{align}
with $F^-(E,\chi) := F^+(-E,\chi)$.\\
\indent The current $J^e$ due to electron-like excitation source terms is calculated \cite{bagwell} using the formula
\begin{align}
J^e(\chi) = & \frac{e}{h} \left(\int^{-\Delta_0}_{-\infty} + \int^{\infty}_{\Delta_0} \right) \frac{1}{|u_0^2 - v_0^2|} \nonumber \\
\; & \times \left[ T^e_{ L \rightarrow R}(E,\chi) - T^e_{ R \rightarrow L}(E,\chi) + \overbar{ T^e_{ L \rightarrow R}}(E,\chi) \right. \nonumber \\
\; & \left. - \overbar{ T^e_{ R \rightarrow L}}(E,\chi) \right] f(E) \mathrm{d} E.
\label{eq:app_Jnl}
\end{align}
Here, $f(E)$ is the Fermi-Dirac distribution at temperature $T$. It can be seen that Eq. \ref{eq:app_Jnl} is equal to Eq. \ref{eq:continuum-current}. In deriving Eq. \ref{eq:app_Jnl} we used only the electrical current transmitted due to electron-like source terms. Repeating the above calculations for hole-like source terms results in a current $J^h$ which is equal to $J^e$. Naively, one might then think the total continuum current obtained is too large by a factor of two. However, note that the density of excitations (electron- plus hole-like) in the $S$-section is twice as large as the density of states in the $N$-section. Consequently, the subband's continuum current is
\begin{equation}
J(\chi) = \frac{1}{2}\left( J^e(\chi) + J^h(\chi) \right),
\end{equation}
and we recover Eq. \ref{eq:continuum-current}. This disparity in the density of states in the $S$- and $N$-sections was noted in Ref. \cite{bagwell}, see Eq. B8 in that paper.\\
\indent Equation \ref{eq:app_Jnl} is used in section \ref{sec:3_results} to numerically calculate the continuum current $J$ due to each subband. For a junction much shorter than the subband's healing length, $L \ll \xi^0$, the bound state current $I$ is much larger than $J$. When $L \gtrsim \xi^0$, $I_{n,l}$ and $J_{n,l}$ are of the same order of magnitude, and we obtain a triangular CPR. At finite magnetic fields, phase shifts appear in both $I$ and $J$ as described in section \ref{sec:31_one_subband}, but the CPR retains its triangular shape.\\
\indent Notice that at zero magnetic field, we have $k^e(-E) = k^h(E)$, so $J$ simplifies and can be written as
\begin{align}
J(\chi) = & \frac{2e}{h} \left(\int^{-\Delta_0}_{-\infty} + \int^{\infty}_{\Delta_0} \right) \frac{1}{|u_0^2 - v_0^2|} \nonumber \\
\; & \times \left[ T^e_{ L \rightarrow R}(E,\chi) - T^e_{ R \rightarrow L}(E,\chi) \right] f(E) \mathrm{d} E.
\end{align}
The extra factor of 2 here is usually attributed to the spin degree of freedom of the electrons/holes. By including both types of coefficients $T^e, \overbar{T^e}$ in Eq. \ref{eq:app_Jnl} we are taking into account this degree of freedom. This can be further elucidated by noting that in the spinful version of this problem, the particle-hole symmetry is manifested as two (Nambu-spinor type) solutions $\bo \Psi$ and $(\sigma_y \tau_y \bo \Psi)^*$ with opposite energies, spins, and coherence factors (the Pauli matrices $\sigma, \tau$ act on the spin and particle-hole manifolds, respectively). That is, $(u, v)^T$ and $(-v^*, u^*)^T$ generalize to states of opposite spin, and the spin degree of freedom is correctly accounted for by considering both types of solutions.

\end{document}